\pdfoutput=1

\documentclass[journal]{IEEEtran}

\usepackage{graphicx} 
\usepackage{subfig}
\usepackage{multirow}

\usepackage{cite}

\usepackage{graphicx}
\usepackage{epstopdf}
\usepackage{textcomp}
\usepackage{xcolor}
\usepackage{verbatim}
\usepackage{makecell}
\usepackage{booktabs}
\usepackage{CJK}
\usepackage{multirow}
\usepackage{bbm}

\usepackage{amsfonts}
\usepackage{amsmath}
\usepackage{dsfont}
\usepackage{amsthm}
\usepackage{amssymb}

\usepackage{algorithm}
\usepackage{multicol} %
\makeatletter
\newcommand{\removelatexerror}{\let\@latex@error\@gobble}
\makeatother

\usepackage{algorithm}
\usepackage{array}
\usepackage{algpseudocode}
\usepackage{amsmath, amssymb}

\begin{document}

\title{Importance-Aware Image Segmentation-based Semantic Communication for Autonomous Driving}
\author{Jie Lv \IEEEauthorrefmark{1}\IEEEauthorrefmark{2},
Haonan Tong \IEEEauthorrefmark{1}\IEEEauthorrefmark{2}, 
Qiang Pan \IEEEauthorrefmark{1}\IEEEauthorrefmark{2}, 
Zhilong Zhang\IEEEauthorrefmark{1}\IEEEauthorrefmark{2}, 
Xinxin He\IEEEauthorrefmark{1}\IEEEauthorrefmark{2}, 
Tao Luo\IEEEauthorrefmark{1}\IEEEauthorrefmark{2}, 
Changchuan Yin \IEEEauthorrefmark{1}\IEEEauthorrefmark{2}
\\
\IEEEauthorrefmark{1}
School of Information and Communication Engineering, Beijing University of Posts and Telecommunications, Beijing 100876, China
\\
\IEEEauthorrefmark{2}
Beijing Key Laboratory of Network System Architecture and Convergence, Beijing 100876, China
\\
Email: \{lvj, hntong\}@bupt.edu.cn
\thanks {Corresponding author:  Tao Luo. Email: tluo@bupt.edu.cn}
}


\maketitle

\begin{abstract}
    This article studies the problem of image segmentation-based semantic communication in autonomous driving.
    In real traffic scenes, detecting the key objects (e.g., vehicles, pedestrians and obstacles) is more crucial than that of other objects to guarantee driving safety.
    Therefore, we propose a vehicular image segmentation-oriented semantic communication system, termed VIS-SemCom, where image segmentation features of important objects are transmitted to reduce transmission redundancy. 
    First, to accurately extract image semantics, we develop a semantic codec based on Swin Transformer architecture, which expands the perceptual field thus improving the segmentation accuracy. 
   Next, we propose a multi-scale semantic extraction scheme via assigning the number of Swin Transformer blocks for diverse resolution features, thus highlighting the important objects' accuracy.
   Furthermore, the importance-aware loss is invoked to emphasize the important objects, and an online hard sample mining~(OHEM) strategy is proposed to handle small sample issues in the dataset.
Experimental results demonstrate that the proposed VIS-SemCom can achieve a coding gain of nearly 6 dB with a 60\% mean intersection over union (mIoU), reduce the transmitted data amount by up to 70\% with a 60\% mIoU, and improve the segmentation intersection over union (IoU) of important objects by 4\%, compared to traditional transmission scheme.
\end{abstract}


\section{Introduction}

Emerging autonomous driving requires vehicles to make intelligent decisions with extremely high perceptual accuracy and low latency~\cite{Chen2020}. To meet the requirements, autonomous vehicles need not only the environmental sensing data from the sensors on themselves, but also the sensing data from adjacent vehicles or roadside infrastructures via vehicle-to-everything (V2X) communications. 
Due to the limited sensing capabilities of a single vehicle, a substantial volume of image data transmissions through V2X are indispensable to facilitate vehicle understanding of traffic scenes, and thus improve driving safety~\cite{Chen2020}, where the detection of objects closely related to safe-driving such as vehicles, pedestrians and obstacles is more important than that of other objects (e.g., sky and fence).
\emph{However, the transmission of massive sensing data will consume large amounts of wireless bandwidth resources, which poses a huge challenge for resources-constrained V2X communication}~\cite{Zhou2020}.
To this end, substantial research efforts have studied higher frequency bands~\cite{Choi2016, Kong2017, Memedi2020, Almarashli2018}, including mmWave, terahertz (THz), and visible light, as well as the large-scale antenna technology to enhance spectrum efficiency for V2X communication, however, the ability to accurately receive transmitted bits is progressively approaching the Shannon capacity limit.
From another perspective, the semantic communication paradigm, focusing on transmitting and receiving the intrinsic meanings at the semantic level, has aroused wide concerns for the high communication efficiency~\cite {Bao2011, Emilio2021, Zhang2022}. 
Concretely, semantic communication selectively extracts a compact set of core data containing valuable information for the target task, and reconstructs the original meanings through semantic inference based on a shared knowledge base.
With the advancements in machine learning for extracting data features, semantic communication has been extensively investigated across diverse domains, including text~\cite{Xie2021, Xie2020, Jiang2022}, speech~\cite{Weng2021, Weng20212, Tong2021}, image~\cite{Bourtsoulatze2019, Kurka2020, Jankowski2020, Xu2022}, video~\cite{Jiang2023, Wang2023, Liang2023}, and more.
These works~\cite{Xie2021, Xie2020, Jiang2022, Weng2021, Weng20212, Tong2021, Bourtsoulatze2019, Kurka2020, Jankowski2020, Xu2022, Jiang2023, Wang2023, Liang2023} have proved that semantic communication can obtain a significant reduction in transmission overhead and enable robust communication even in poor channel conditions. 
Given the powerful computing capabilities of intelligent vehicles, utilizing semantic communication for transmitting extensive vehicle image data has become a potential solution for communication-efficient V2X image transmission.

The existing literature has studied semantic communication in the context of V2X~\cite{Chen2022, Xu2023, Xia2023, Su2023}. 
The work in~\cite{Chen2022} proposed a resource allocation scheme to maximize the semantic understanding accuracy of intelligent tasks with the least resource cost, while the differentiated demand of vehicle users and adaptive resource allocation schemes have not been considered.
Furthermore, the study in~\cite{Xu2023} designed a Cooperative Semantic Communication (Co-SC) architecture tailored for multi-user scenarios, so as to eliminate redundancy in transmitted data among multiple users.
Through the integration of a cooperative semantic decoder and the joint optimization of semantic encoders and the cooperative semantic decoder, the Co-SC architecture was capable of learning semantic-level correlations among users, thus significantly reducing the data transmission.
Besides, considering the challenge of knowledge base matching for multiple users in Vehicle-to-Vehicle (V2V) networks, the authors in~\cite{Xia2023} jointly addressed the knowledge base construction (KBC) and vehicle service pairing (VSP) by a novel semantic communication-empowered service supplying solution with low computational complexity. 
The authors in~\cite{Su2023} proposed a long-term robust resource allocation scheme by considering semantic access control, power control, and device-to-device vehicular communication to ensure efficient and reliable communications in dynamic environments. 
Nonetheless, these works~\cite{Chen2022, Xu2023, Xia2023, Su2023} primarily concentrated on resource allocation schemes for transmitting semantic data rather than semantic codec design, which is crucial for further enhancing the efficiency of V2X image transmission. 

In the domain of image semantic codec design, existing research has predominantly focused on image recovery or classification tasks at the receiver~\cite{Bourtsoulatze2019, Kurka2020, Jankowski2020, Xu2022},
in which images were compressed via understanding image contents based on neural networks~(NNs). 
In~\cite{Bourtsoulatze2019}, the authors proposed a deep joint source-channel coding (JSCC) scheme to map image pixels directly to the channel input and the decoder uses NN to reconstruct the image, which significantly improved the reconstruction quality under low signal-to-noise ratios (SNR).  
In~\cite{Hu2022}, a masked autoencoder (MAE) based on  Vision Transformer (ViT) architecture was proposed to alleviate semantic noise on patches of images and even restore the missing patches, which proved that ViT achieved a more accurate perception capacity.
Moreover, the authors in~\cite{yoo2023role} introduced a novel measure to analyze the fundamental functions of different CNN variants and proposed a ViT-based mode in semantic communication, which can achieve a peak signal-to-noise ratio (PSNR) gain of 0.5 dB. 
The work in~\cite{Yang2023} further proved that the Swin Transformer-based Deep JSCC model can save more channel bandwidth under various channel conditions for high-resolution images. 
However, these works~\cite{Bourtsoulatze2019, Kurka2020, Jankowski2020, Xu2022} still had redundant data transmission without considering that autonomous driving necessitates the accurate understanding of important objects~\cite{Chen2019} that can be derived from small-sized semantic segmentation data. 
Thus, how to construct semantic segmentation-oriented network architecture to meet the demands of important objects for safe autonomous driving is worthwhile being explored.

In this article, we extend our prior work in~\cite{Pan2023} and propose a vehicular image segmentation-oriented semantic communication (VIS-SemCom) system, in which image segmentation features are transmitted and the segmentation results are recovered directly at the receiver. 
In VIS-SemCom, Swin Transform-based codecs are designed to extract multi-scale semantics of images. 
Meanwhile, the codec structure and loss function are carefully designed to improve the segmentation accuracy of important objects. To the best of our knowledge, \emph{this is the first work that designs vehicular semantic communication considering the object importance in images to improve transmission performance.}

The main contributions can be summarized as follows:
\begin{itemize}
	\item In light of that the autonomous driving scene should pay more attention to understanding important objects such as vehicles, pedestrians, and obstacles on the road, we propose a semantic communication system for the vehicular image segmentation task. Unlike existing work of original image recovery with equal object importance, the transceivers perform image segmentation to obtain semantic label maps. To harness the full potential of semantic communications to save transmission resources, we extract multi-scale semantic features based on the input traffic image and transmit only small amounts of extracted semantic feature information through a wireless channel.
	
	\item Inspired by advanced computer vision technology, we design a Swin Transformer-based encoder to extract multi-scale semantic features. Specifically, a multi-level local self-attention mechanism and a shift window strategy are leveraged to balance the semantic extraction capability and computational complexity of the neural network. To emphasize accurate segmentation of important objects, the system including network structure, loss function, and training strategy is optimized. We propose an importance-aware loss function jointly considering the weighted multi-class cross-entropy and intersection over union (IoU). Additionally, an online hard sample mining (OHME) training strategy is utilized to deal with relatively small samples in the data set. 
	
	\item We consider the image segmentation accuracy IoU as a performance metric. Simulation results demonstrated that, compared to traditional image transmission schemes without semantic extraction operations, the proposed VIS-SemCom system can significantly reduce data volume by approximately 70\% and obtain a coding gain of 6 dB when achieving a segmentation mIoU of 60\%. Besides, the IoU of important objects can be improved by about $4\%$.
\end{itemize}

The rest of this paper is organized as follows. In Section II, we first describe the system model and highlight the structure of the proposed VIS-SemCom. The details of the proposed VIS-SemCom are presented in Section III. Section IV shows the simulation results, and Section V concludes this paper.

\section{System Model}

\begin{figure*}[t]
	\centering
	\subfloat[A scenario of V2V image transmission.]{
		\includegraphics[width = 0.48\textwidth]{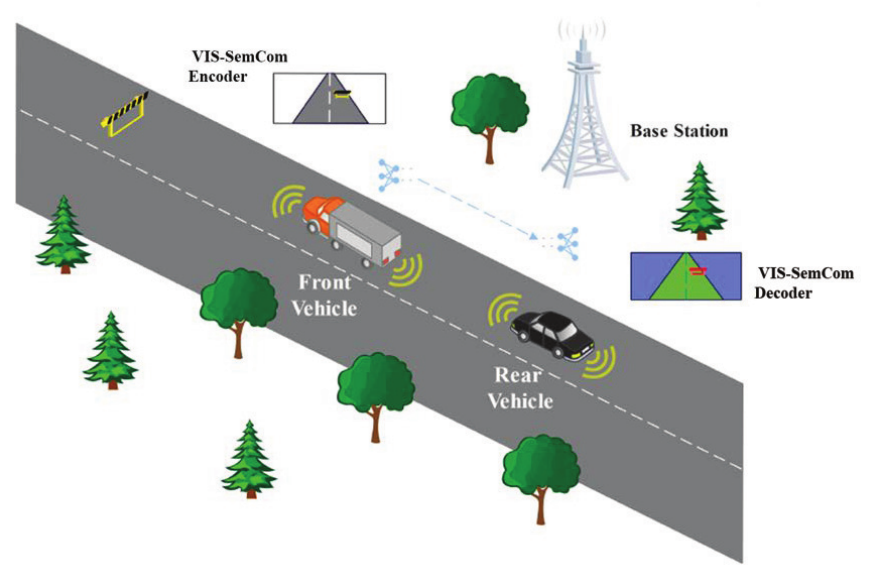}\label{scenario}
	}
	\subfloat[The structure of the poposed VIS-SemCom.]{
		\includegraphics[width = 0.48\textwidth]{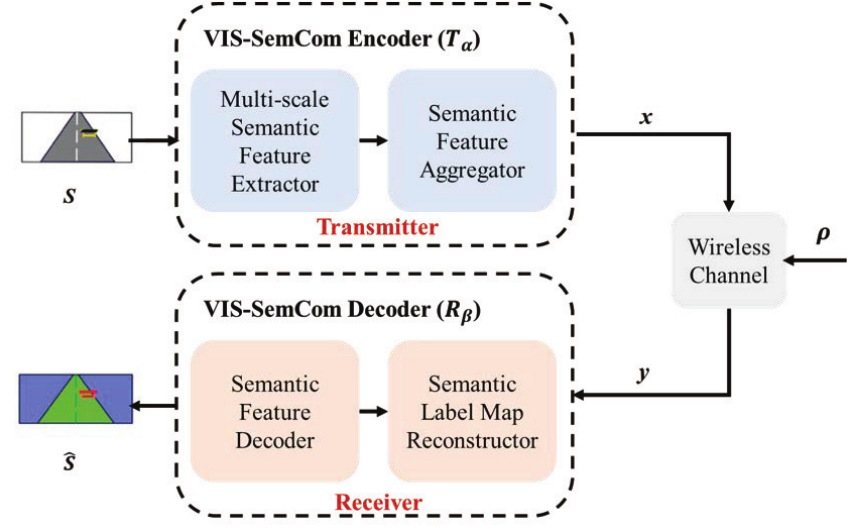}\label{structure}
	}
	\caption{The VIS-SemCom system model.}
	\label{system_model}	
\end{figure*}

We consider a scenario of Vehicle-to-Vehicle (V2V) image transmission to extend the perceptual range and enhance traffic scene understanding. As shown in Figure~\ref{system_model}\subref{scenario}, two vehicles are traveling in the same direction and one car is behind a truck. When there is an obstacle in front of the front orange truck, the truck camera can detect the situation, but it is beyond the field of view of the rear black car. To avoid collision and make wise decisions for lane change, massive sensor images should be shared with the car by V2V communications and an accurate segmentation of important objects should be obtained at the receiving vehicle. Hence, we propose the VIS-SemCom system mounted on vehicles. 

The overall structure of the VIS-SemCom system is shown in Figure~\ref{system_model}\subref{structure}, which includes two components. (\textit{i}) the VIS-SemCom encoder at the transmitter (e.g., the front truck), which extracts semantic features of input images for transmission; (\textit{ii}) the VIS-SemCom decoder at the receiver (e.g., the rear car), which decodes received semantic features and completes the reconstruction of semantic label maps with the category and location information of the object. 
Next, we first introduce the semantic encoder module and the decoder module, followed by the formulation of performance metrics.

\subsection{Semantic Encoder Module}
\label{s3-1}

At the transmitter, an input image is represented by a vector $\boldsymbol{S}\in \mathbb{R}^{H \times W \times 3}$, where $H$, $W$, $3$ denotes the height, weight, and the number of channels, respectively. The VIS-SemCom semantic encoder extracts and aggregates $\boldsymbol{S}$ through a multi-scale semantic feature extractor and a semantic feature aggregator, outputting encoded semantic features to a $k$-length vector of complex-valued,  $\boldsymbol{x} \in \mathbb{C}^k$. This encoder can be parameterized by a function $\boldsymbol{{T_\alpha }}(\cdot)$ with parameters $\boldsymbol{\alpha}$, and the encoding process can be expressed as 
\begin{equation}
\boldsymbol{x} = \boldsymbol{{T_\alpha }}(\boldsymbol{S}).\label{eq1}
\end{equation}

The input image is in RGB format, with pixel values ranging from $[0, 255]$. 
To extract semantic features from shallow to deep, the processed image passes through a multi-scale semantic feature extractor composed of a neural network (NN), and then the semantic feature aggregator fuses multi-scale semantic features as channel input symbols. As a result, the image is compressed and the compression ratio $R$ is defined as the space occupied before compression divided by the actual space occupied, which reflects the quality of the image after compression.  

Next, encoded semantic features are transmitted over the wireless channels, including channel fading and noise, and then the semantic features $\boldsymbol{y}$ received by the VIS-SemCom decoder can be presented as 
\begin{equation}
\boldsymbol{y} =\boldsymbol{h}\boldsymbol{x} + \boldsymbol{\rho},\label{eq2}
\end{equation}
where $\boldsymbol{\rho} \sim \mathcal{CN}\left(0, \sigma^2 \textbf{I} \right)$ is a complex Gaussian noise with mean zero and variance $\sigma^2$, and $\boldsymbol{h}$ is the channel gain. Since there is relative motion between vehicles, the Doppler spread in the channel is introduced and the channel fades become temporally correlated. The time-varying channel gain can be modeled as
\begin{equation}
\boldsymbol{h}(t) = \gamma {e^{j 2\pi {f_d} t}},\label{eq3}
\end{equation}
where $\gamma$ is the channel fading coefficient, and ${f_d}$ is the maximum Doppler frequency shift, which can be calculated by $f_d = f_c v/c$, where $v$ is the vehicle velocity, $f_c$ is the carrier frequency and $c$ is the speed of light.

\subsection{Semantic Decoder Module}

At the receiver, the received semantic feature ${\bf{y}}$ is decoded by a semantic feature decoder, and then reconstructed into a semantic label map ${\bf{\hat S}}$ by a reconstructor. These two parts make up the VIS-SemCom decoder to perform image segmentation tasks. Similarly to the encoding function, the decoding function is represented as $\boldsymbol{R_{\beta}}(\cdot)$ with parameters $\beta$. Then, the results of image segmentation can be obtained by
\begin{equation}
\boldsymbol{\widehat{S}}=\boldsymbol{R_{\beta}}(\boldsymbol{y}).\label{eq4}
\end{equation}

Note that the acquired semantic segmentation $\boldsymbol{\widehat{S}}$, is with the same size (length and width) as the input image. $\boldsymbol{\widehat{S}}$ classifies pixels in the input image $\boldsymbol{S}$ into ${N_{cls}}$ classes, where each value represents the count of object classes in the images. These classes may include vehicles, pedestrians, and so on.

\begin{figure*}[t]
	\centering{\includegraphics[width = 0.9\textwidth]{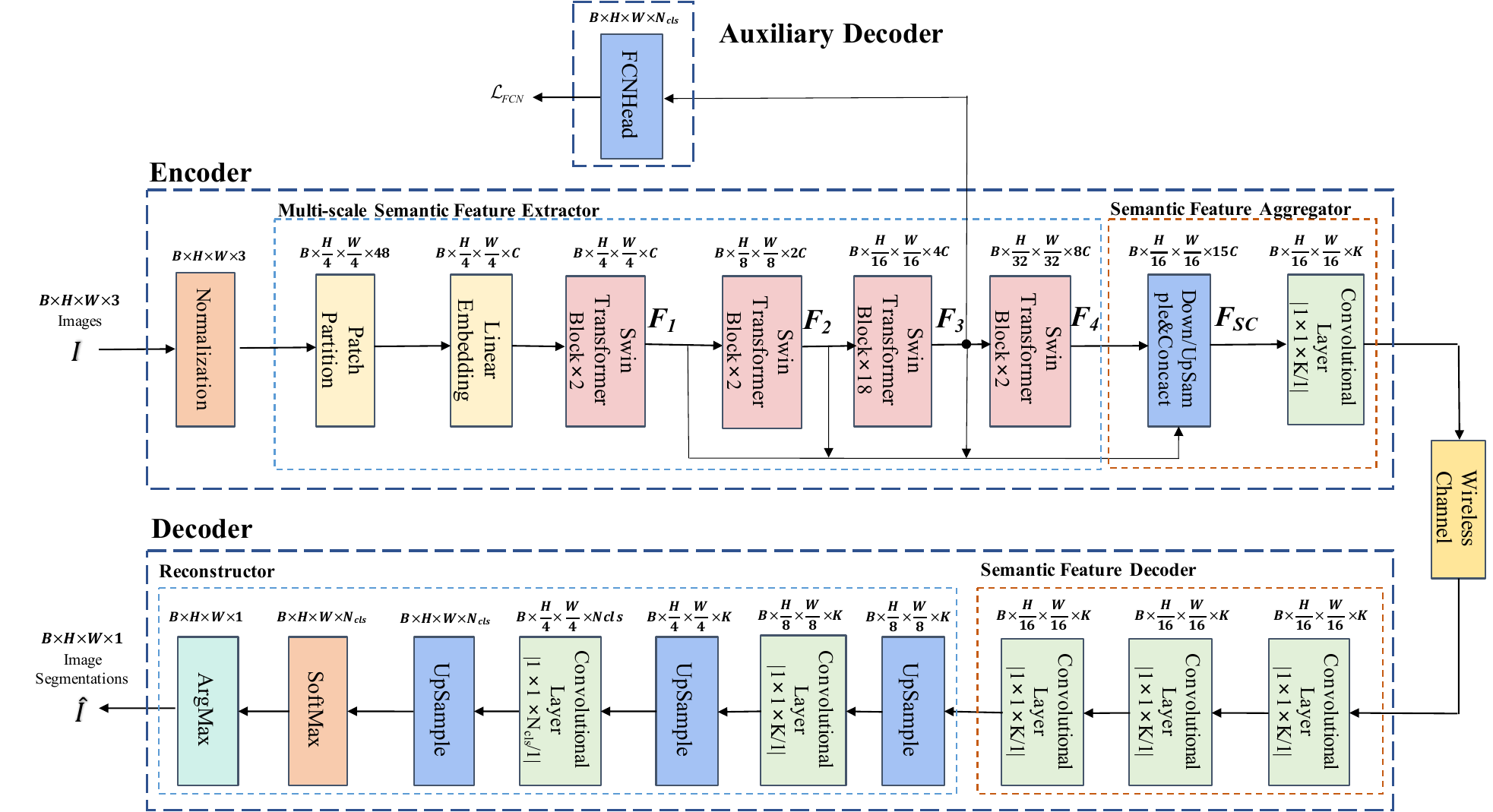}}
	\caption{The framework of the proposed VIS-SemCom system consisting of an encoder and a decoder.}
	\label{network}
\end{figure*}

\subsection{Performance Metrics}
The proposed VIS-SemCom system aims to complete image semantic segmentation tasks under autonomous driving scene understanding, which is only interested in highly relevant objects. 
Therefore, we use the IoU metric to measure the image recovery performance at this semantic level instead of the bit error rate in traditional communication. The IoU measures the ratio of the overlap between predicted segmentation and labels to their union area, which is defined as follows
\begin{equation}\label{eq51}
\mathrm{I o U}_i= \frac{P_i \bigcap G_i}{P_i \bigcup G_i}
\end{equation}
where ${P_i}$ is the set of pixel regions predicted for the class $i$ object, and ${G_i}$ is the set of actual pixel regions for the class $i$ object. To further evaluate the segmentation performance of the whole image, the mean intersection over union (mIoU) is employed, which is the average of the ratio of the intersection and union of all classes $N_{cls}$, 
\begin{equation}\label{eq5}
\mathrm{m I o U}=\frac{1}{N_{c l s}} \sum_{i=1}^{N_{c l s}} \frac{P_i \bigcap G_i}{P_i \bigcup G_i}
\end{equation}
Obviously, the higher IoU or mIoU represents the better segmentation performance.

\section{VIS-SemCom Framework for Image Segmentation}

In this section, we propose an image segmentation-oriented transmission framework, VIS-SemCom, as shown in Figure~\ref{network}. Due to the limits of the size of the convolution kernel, the representation capacity and semantic extraction ability of previous image semantic communication systems based on CNN and ViT are limited. Therefore, we exploit Swin Transformer as a new backbone for the VIS-SemCom system to extract multi-scale semantic features. 
The input image is divided into multiple small pieces of the same size and processed by multi-layer Transformer modules successively. 
Each Transformer module uses a window-based local attention mechanism that only focuses on information interactions between adjacent blocks, reducing computational and memory complexity~\cite{Liu2021}. 
Next, we introduce the network structure, loss function design, and training strategy.

\subsection{Network Structure}
To transmit the semantic features of important objects through a wireless channel, we first design the network structures of the multi-scale semantic feature extractor and aggregator, then the decoder and reconstructor.

\subsubsection{Multi-scale Semantic Feature Extractor and Aggregator}
As depicted in Figure~\ref{network}, an image undergoes normalization before entering the multi-scale semantic feature extractor. 
This normalization step aims to expedite convergence speed and minimize calculation loss.
The multi-scale semantic feature comprises a patch partition module, a liner embedding layer, and four semantic feature extraction stages. Each semantic feature extraction stage consists of an even number of Swin Transformer blocks (STBs). The semantic feature aggregator consists of an up/down sampling layer and a convolutional layer. 

To intuitively explain the role of each module, we take an input $\boldsymbol{I}\in {{\mathbb{R}}^{B\times H\times W\times 3}}$ as an example, which represents a batch RGB three-channel image, where $B$ is the batch size, and the RGB value is in [0, 255]. The pixel value of the image is normalized to [0, 1] by the normalization layer. The patch partition module divides each $H\times W\times 3$-sized image into several small patches with the size of $4\times4\times3$, and the liner embedding layer converts the channel length of each image block to the size of $C$. 

\begin{figure}[t]  
	\centering
	\subfloat[]{
		\includegraphics[width=0.2\textwidth]{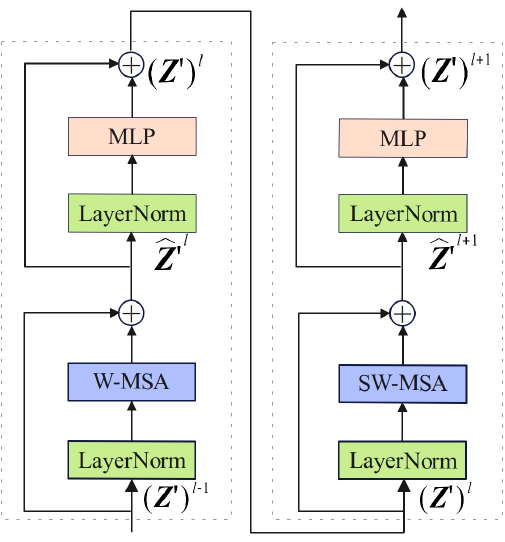}
		\label{STB}
	}
	\subfloat[]{
		\includegraphics[width=0.2\textwidth]{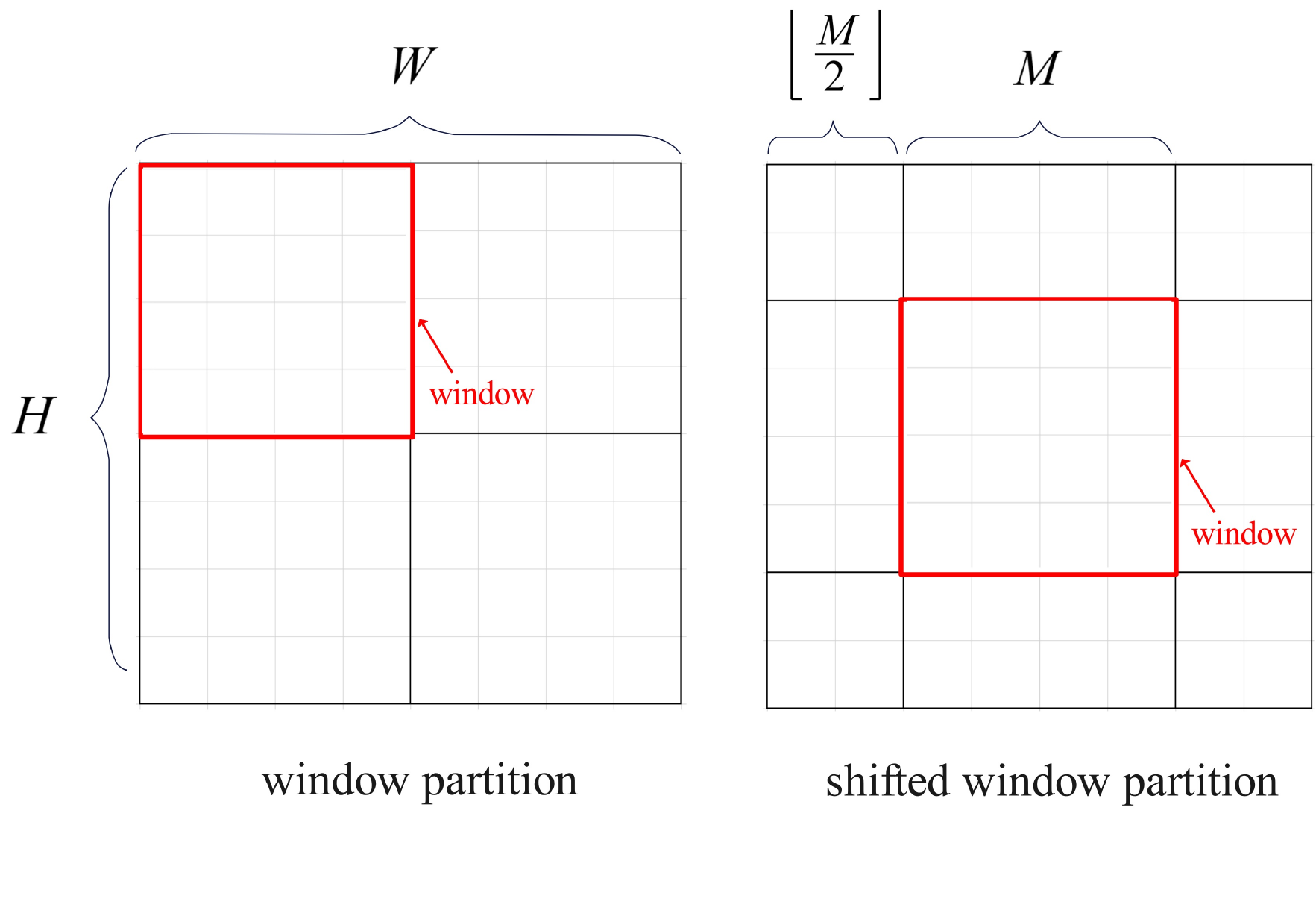}
		\label{window} 
	} 
	\centering
	\caption{(a) The architecture of two consecutive STBs; (b) Illustration of the window partition and shifted window partition (e.g., \textit{H} = 8, \textit{W} = 8, and \textit{M} = 4).}
	\label{STB&MLP}
\end{figure}

Then, these image blocks are fed into four semantic feature extraction stages with the architecture of STBs. As shown in Figure~\ref{STB&MLP}\subref{STB}, an STB is built by a multi-head self-attention (MSA) module based on windows structure. Specifically, an STB divides the input image into many independent small regions, called window partition, as shown in Figure~\ref{STB&MLP}\subref{window} left, and computes self-attention by the window-based MSA (W-MSA) module separately. To provide interaction among adjacent pixels in different windows, the shifted window MSA (SW-MSA) module is used between consecutive STBs, significantly extending the perceptual field. Besides, a multi-layer perceptron (MLP) layer and two LayerNorm (LN) layers are contained, which together with the (S)W-MSA module form two residual modules.
For the input $\boldsymbol{Z'}$ by window partition, the consecutive STBs $l$ and $l+1$ of a stage are given by
\begin{equation}   
\begin{aligned}
&\boldsymbol{\widehat{Z'}}^{l}={\rm W\mbox{-}MSA(LN}((\boldsymbol{Z'})^{l-1})) + (\boldsymbol{Z'})^{l-1}, \\       
&(\boldsymbol{Z'})^{l}={\rm MLP(LN}(\boldsymbol{{\widehat{Z'}}}^{l})) + \boldsymbol{{\widehat{Z'}}}^{l},   \\ \label{process1}
&\boldsymbol{\widehat{Z'}}^{l+1}={\rm SW\mbox{-}MSA(LN}((\boldsymbol{Z'})^{l})) + (\boldsymbol{Z'})^{l}, \\       
&(\boldsymbol{Z'})^{l+1}={\rm MLP(LN}(\boldsymbol{{\widehat{Z'}}}^{l+1})) + \boldsymbol{{\widehat{Z'}}}^{l+1},  
\end{aligned}
\end{equation}
where $\boldsymbol{\widehat{Z'}}^{l}$ and $(\boldsymbol{Z'})^{l}$ denotes the output feature of the (S)W-MSA module and the MLP module for block ${l}$, respectively.   

Before entering the next stage of the semantic feature extraction layer, the length and width of the image become half of the original size, and the number of channels becomes twice. This process realizes the extraction of image semantic information from a high-resolution feature map to a low-resolution feature map, corresponding to  
the semantic feature maps ${{F}_{1}}$, ${{F}_{2}}$, ${{F}_{3}}$ and ${{F}_{4}}$, respectively. In turn, richer semantic content is possessed. This is because the high-resolution feature map can only see the local part of the image, focusing on the edge information, and the low-resolution feature map can see the global image. As a result, high-resolution feature maps are mainly used for image classification tasks, while image segmentation and object detection tasks are more affected by low-resolution feature maps.
For example, in the classical U-net model structure~\cite{Ronneberger2015}, the low-resolution feature map is sampled to the same size as the high-resolution feature map, and the fusion is performed by far-jump connection, improving the segmentation accuracy of the image. In our proposed network structure, different values are assigned for STBs with different resolutions to enhance the semantic extraction of objects with differentiated segmentation requirements, especially, the semantic extraction of low-resolution feature maps is increased to improve the segmentation accuracy of important objects usually with large sizes. 

In the semantic feature aggregator, the VIS-SemCom system samples the length and width of ${{F}_{1}}$, ${{F}_{2}}$ and ${{F}_{4}}$ to the size of ${{F}_{3}}$, and the sampling method is a bilinear interpolation. 
When the length and width of four semantic feature maps are the same, they are spliced in the channel direction to obtain the semantic feature map ${{F}_{sc}}$. Finally, ${{F}_{sc}}$ is fed into the convolution layer, which has $K$ convolutions with $1\times 1$ convolution kernels to compress data. 
The value of $K$ is used to adjust the compression ratio, and a smaller $K$ denotes a larger compression ratio.

\subsubsection{Semantic Feature Decoder and Reconstructor}
The semantic feature decoder contains three convolutional layers, which are used to reduce the influence of channel noise on the semantic features of the receiver. Each convolution layer has $K$  filters of size 1 with a step size of 1. 
The reconstructor consists of three upsampling layers, two convolution layers, a softmax activation function layer and an argmax layer. 
The upsampling layer is employed to progressively restore the length and width of the original image, while the convolutional layer is used to sequentially extract semantic information.
The first convolutional layer has $K$ filters of size 1 with a step size of 1, and the second convolutional layer has $N_{cls}$ filters of size 1 with a step size of 1. 
The three upsampling layers are arranged in sequence with the two convolutional layers, and the length and width of the semantic features are gradually changed to the same as the input. 
Next, the semantic feature map $P\in {{\mathbb{R}}^{B\times H\times W\times {{N}_{cls}}}}$ is obtained by a softmax activation function, where the value of each internal element is the predicted value of pixel for multi-classification.
Finally, performing argmax operation, the semantic feature graph $P$ is converted to the semantic label map $\boldsymbol{\widehat{I}}\in {{\mathbb{R}}^{B\times H\times W\times 1}}$, and the value inside the semantic label map is the index value of the predicted pixel class.

\subsection{Training Loss Function}

As mentioned in the introduction, driving safety is of paramount importance, with collisions involving cars, pedestrians, and bicyclists being particularly critical.
In contrast, the sky and the fence are considered less critical as they are not directly related to road conditions.
Since different object classes have varying degrees of importance in autonomous driving, we propose an importance-aware loss function that combines the weighted multi-class cross-entropy with the Intersection over Union (IoU) loss for crucial objects.

First, the weighted multi-class cross-entropy is given by 
\begin{equation}
{{\cal L}^{\mathrm{w}}_{CE}} = - \sum\limits_{i = 1}^{N_{cls}} {({w_i}{y_i}\log (} {p_i})),
\end{equation}
where ${p_i}$ is the possibility that a pixel is classified into object class $i$, and ${y_i}$ is the indicator of whether the pixel belongs to the object;
and ${w_i}$ is the weight coefficient obtained by multiplying the category balance loss coefficient by the attention coefficient. 
The category balance loss coefficient is set based on~\cite{Chen2017}, and the attention coefficient is determined as 1 or 1.5, depending on its importance.
The weight coefficient puts more emphasis on accurately segmenting important objects than less important ones. 
For a batch of $B$ images, the multi-class cross-entropy can be given by \begin{equation}
{\cal L}_{W} = {\textstyle{{\sum\limits_{n = 1}^B {\sum\limits_{m = 1}^{H \times W} {{\cal L}^{\mathrm{w}}_{CE}} } } \over {B \times H \times W}}}.
\label{WCE}
\end{equation}


To mitigate semantic errors caused by data imbalance, the IoU loss function of the important objects is expressed as
\begin{equation}
{{\cal L}_{IoU}} = {{\sum\limits_{i = 1}^{{N_{I}}} {1 - {{{P_i} \cap {G_i}} \over {{P_i} \cup {G_i}}}} } \over {{N_{I}}}},
\label{IoU_loss}
\end{equation}
where ${N_{I}}$ is the number of objects with high importance.

By combining \eqref{WCE}-\eqref{IoU_loss} the importance-aware loss function of the VIS-SemCom system is obtained by 
\begin{equation}
{{\cal L}_{IA}} ={{\cal L}_{IoU}} + {{\cal L}_{W}}.
\label{IA_loss}
\end{equation}
In addition, because the semantic feature extractor directly determines the quality of the semantic label map reconstructed by the receiving end, an auxiliary decoder composed of full convolutional network (FCN) is added in the training stage shown in Figure~\ref{network}.  
Its internal is composed of transposed convolution, and the auxiliary decoder also uses the importance-aware loss function, denoted by ${{\cal L}_{F}}$.
The auxiliary solution dock plays a role in jumping out of the local minimum value and preventing over-fitting in training. Therefore, the final training loss of the VIS-SemCom system is 
\begin{equation}
{{\cal L} = {b_1}{{\cal L}_{IA}} + {b_2}{{\cal L}_{F}}},
\label{total_loss}
\end{equation}
where ${b_1}$ is set to 1 and ${b_2}$ is set to 0.4, which is the recommended configuration of mmSegmentation framework~\cite{mmS2020}.

\subsection{Online Hard Sample Mining Strategy}

\label{training_strategy}

Given the above loss function, the VIS-SemCom system is trained in an end-to-end manner, which can learn the characteristics of the channel and reduce the impact of channel noise during decoding.
However, a set of important objects (such as rider, bicycle, motorcycle, etc.) constitutes a small proportion of the dataset. This presents a challenge for the model to effectively learn features from these instances, commonly referred to as hard samples. 
Therefore, the OHEM training strategy is adopted to improve the learning efficiency of small samples. 

The workflow of the OHEM strategy is outlined in Algorithm~\ref{algorithm1}, primarily involving two parameter settings: $thresh$ and $min\_kept$. The $thresh$ parameter represents the confidence score, and only pixels with confidence scores below the threshold will undergo training.
The parameter $min\_kept$ represents the minimum number of pixels that must be retained in the training batch, which can consolidate the training results to a certain extent, so that the forward propagation will not be wasted.

\section{Simulation Results}

\subsection{Settings and Baselines}

\begin{table}[t]
	\centering
	\caption{Simulation parameters}
	\resizebox{\linewidth}{!}{
    \fontsize{40}{60}\selectfont
		\begin{tabular}{|c|c|c|c|}
			\hline
			\multicolumn{4}{|c|}{\textbf{VIS-SemCom system network parameter settings}} \\
			\hline
			\multicolumn{1}{|c|}{\textbf{Module}} & \textbf{Layer Name} & \multicolumn{1}{c|}{\textbf{Parameters}} & \textbf{Values} \\
			\hline
			\multicolumn{1}{|c|}{\multirow{7}[14]{*}{\makecell[c]{Multi-scale \\ Semantic Feature\\ Extractor}}} & \multicolumn{1}{c|}{\multirow{7}[14]{*}{\makecell[c]{4${\times}$Swin \\ Transformer\\ Stage}}} & \multicolumn{1}{c|}{depths} & [2, 2, 18, 2] \\
			\cline{3-4}    \multicolumn{1}{|c|}{} &       & \multicolumn{1}{c|}{head number} & [3, 6, 12, 14] \\
			\cline{3-4}    \multicolumn{1}{|c|}{} &       & \multicolumn{1}{c|}{\makecell[c]{embedding \\ dimension}}  & 96 \\
			\cline{3-4}    \multicolumn{1}{|c|}{} &       & \multicolumn{1}{c|}{window size} & 7 \\
			\cline{3-4}    \multicolumn{1}{|c|}{} &       & \multicolumn{1}{c|}{patch size} & 4 \\
			\cline{3-4}    \multicolumn{1}{|c|}{} &       & \multicolumn{1}{c|}{mlp ratio} & 4 \\
			\cline{3-4}    \multicolumn{1}{|c|}{} &       & \multicolumn{1}{c|}{activation} & GELU \\
			\hline
			\multicolumn{1}{|c|}{\multirow{4}[8]{*}{\makecell[c]{\makecell[c]{Semantic Feature\\ Aggregator}}}} & {\makecell[c]{4${\times}$down/up \\ sampling layer}}  & \multicolumn{1}{c|}{sampling ratio} & [1/4, 1/2, 1, 2] \\
			\cline{2-4}    \multicolumn{1}{|c|}{} & \multirow{3}[6]{*}{Convolution Layer} & \multicolumn{1}{c|}{kernel size} & 1${\times}$1 \\
			\cline{3-4}    \multicolumn{1}{|c|}{} &       & \multicolumn{1}{c|}{stride} & 1 \\
			\cline{3-4}    \multicolumn{1}{|c|}{} &       & \multicolumn{1}{c|}{number} & \textit{K} \\
			\hline
			\multicolumn{1}{|c|}{\multirow{3}[6]{*}{\makecell[c]{Semantic Feature \\ Decoder}}} & \multirow{3}[6]{*}{3${\times}$Convolution Layer} & \multicolumn{1}{c|}{kernel size} & 1${\times}$1 \\
			\cline{3-4}    \multicolumn{1}{|c|}{} &       & \multicolumn{1}{c|}{stride} & 1 \\
			\cline{3-4}    \multicolumn{1}{|c|}{} &       & \multicolumn{1}{c|}{number} & \textit{K, K, K} \\
			\hline
			\multicolumn{1}{|c|}{\multirow{4}[8]{*}{Reconstructor}} & 3${\times}$ up sampling layer & \multicolumn{1}{c|}{ratio} & 2, 2, 4 \\
			\cline{2-4}    \multicolumn{1}{|c|}{} & \multirow{3}[6]{*}{2${\times}$Convolution Layer} & \multicolumn{1}{c|}{kernel size} & 1${\times}$1 \\
			\cline{3-4}    \multicolumn{1}{|c|}{} &       & \multicolumn{1}{c|}{stride} & 1 \\
			\cline{3-4}    \multicolumn{1}{|c|}{} &       & \multicolumn{1}{c|}{number} & \textit{K, Ncls} \\
			\hline
			\multicolumn{4}{|c|}{\textbf{Physical layer parameter settings}} \\
			\hline
			\multicolumn{2}{|c|}{\textbf{Parameters}} & \multicolumn{2}{c|}{\textbf{Values}} \\
			\hline
			\multicolumn{2}{|c|}{Vehicle Transmitting Power} & \multicolumn{2}{c|}{23 dBm} \\
			\hline
			\multicolumn{2}{|c|}{Vehicle Velocity} & \multicolumn{2}{c|}{50 km/h, 120 km/h} \\
			\hline
			\multicolumn{2}{|c|}{Carrier Frequency} & \multicolumn{2}{c|}{5.9 GHz} \\
			\hline
			\multicolumn{2}{|c|}{Channel Bandwidth} & \multicolumn{2}{c|}{20 MHz} \\
			\hline
			\multicolumn{2}{|c|}{Channel Fading Model} & \multicolumn{2}{c|}{\makecell[c]{Log-Normal shadowing \\ distribution with standard deviation \\ of 8 dB + Rayleigh fading channel} } \\
			\hline
		\end{tabular}
  }
	\label{table1}%
\end{table}%

We train and evaluate the performance of the proposed VIS-SemCom system on the public urban traffic scenario dataset Cityscapes~\cite{Cordts2016}, which is a public semantic segmentation dataset for autonomous driving. 
The dataset consists of 2975 training images, 500 validation and 1525 test images, where the images are in RGB format with $2048\times 1024$ pixels. 
Semantic classification of objects (i.e., the objects in the semantic label map) is divided into 19 classes, where 12 meaningful object classes for driving safety are focused on in the following simulations. The data is further augmented by random cropping, random flipping and photometric distortion.
The transmission power is 23 dBm and bandwidth is 20 MHz. To adapt to the dynamic channel conditions, the VIS-SemCom method is trained under a random distribution within the channel $ SNR_{train} \in [1, 20]$ dB, and Adam optimizer is employed with a learning rate of $1 \times 10^{-4}$. The other parameters of VIS-SemCom network setup are listed in Table~\ref{table1}.

\begin{figure*} [t]
	\centering
	\subfloat[Conventional 19 dB]{
		\begin{minipage}[b]{0.2\textwidth}
			\includegraphics[width=1.5in]{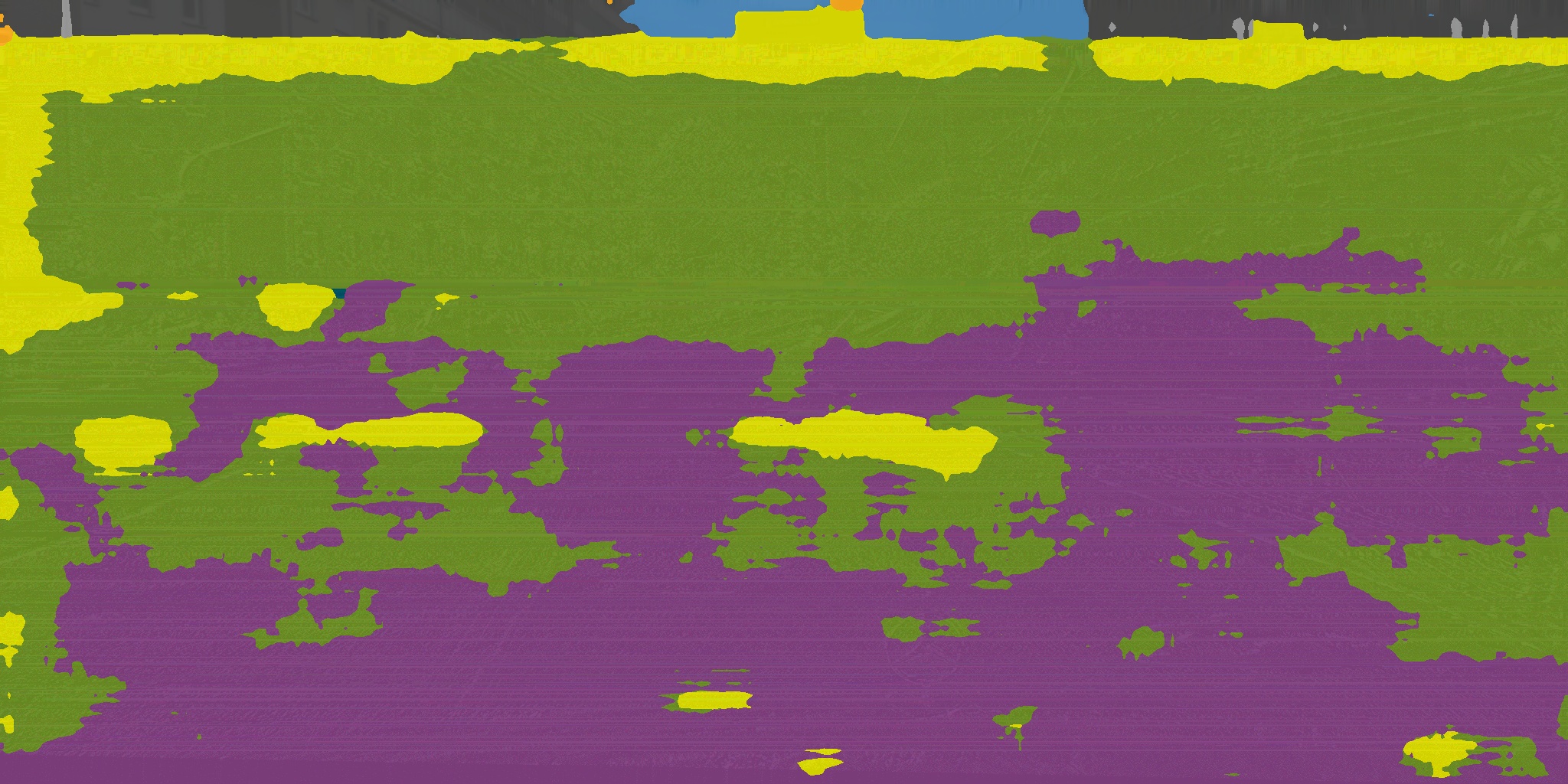}\\ \vspace{-0.4cm}
			\includegraphics[width=1.5in]{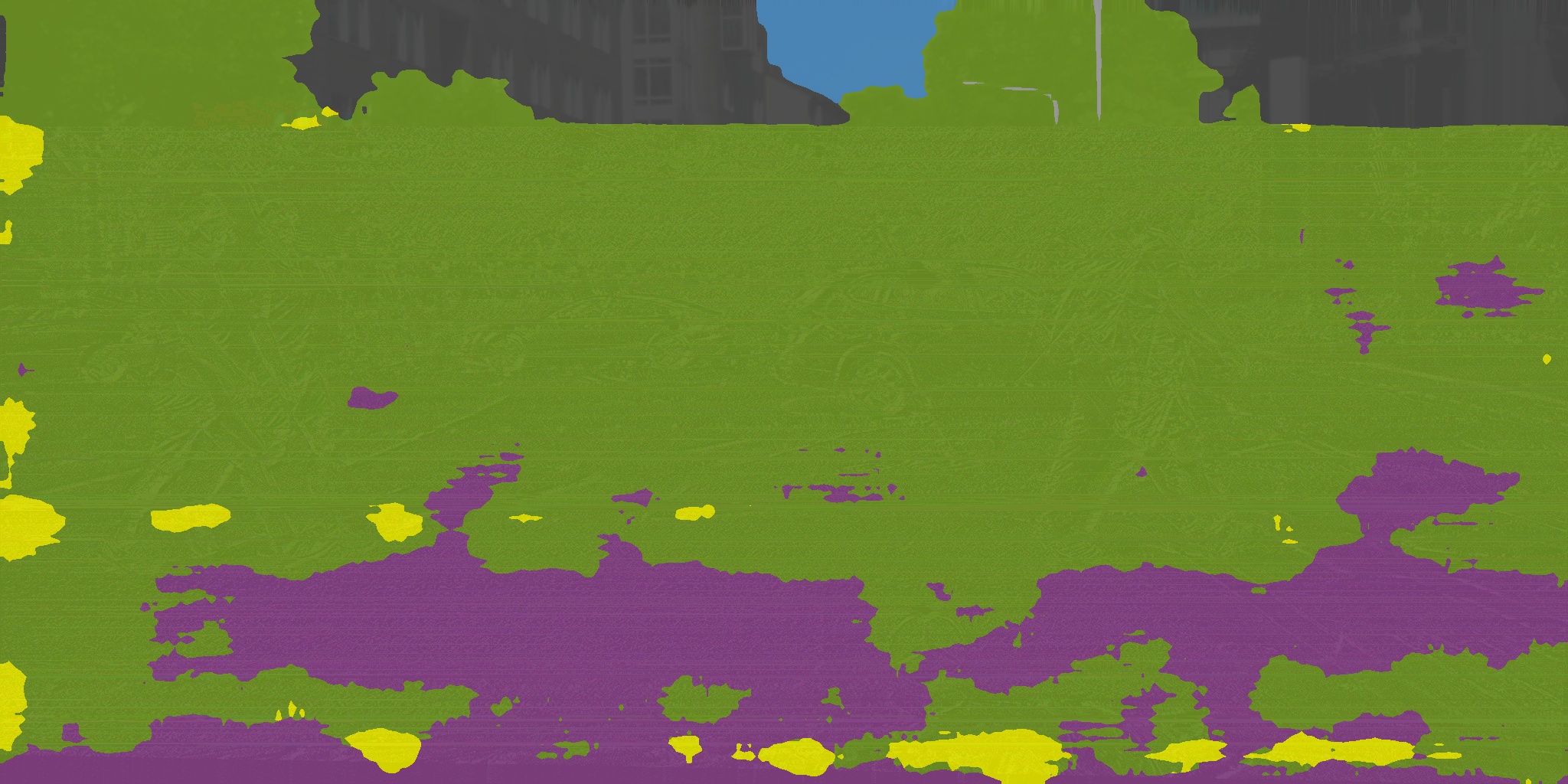}\\
			\vspace{-0.4cm}
			\includegraphics[width=1.5in]{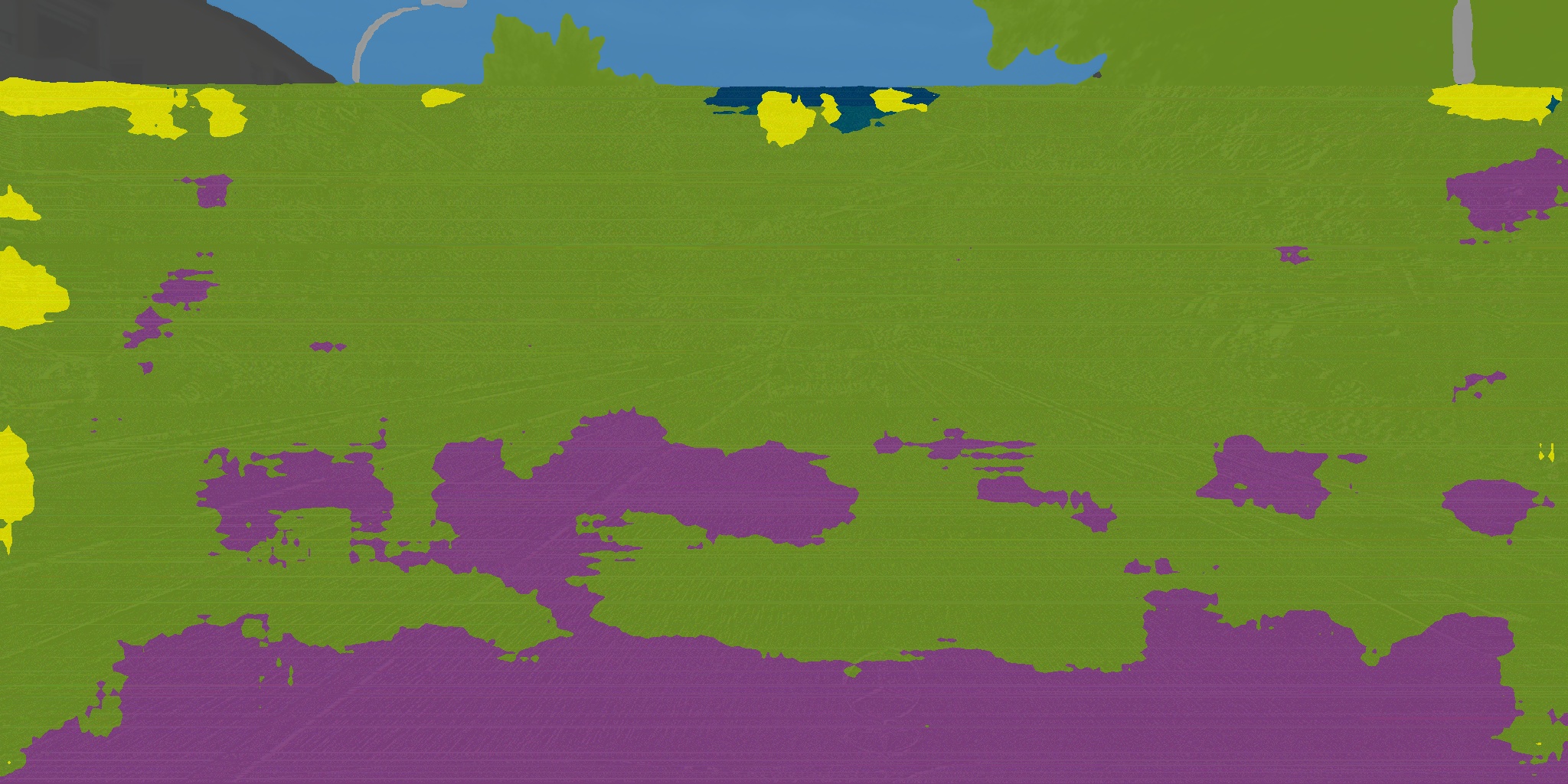}\\
			\vspace{-0.4cm}
			\includegraphics[width=1.5in]{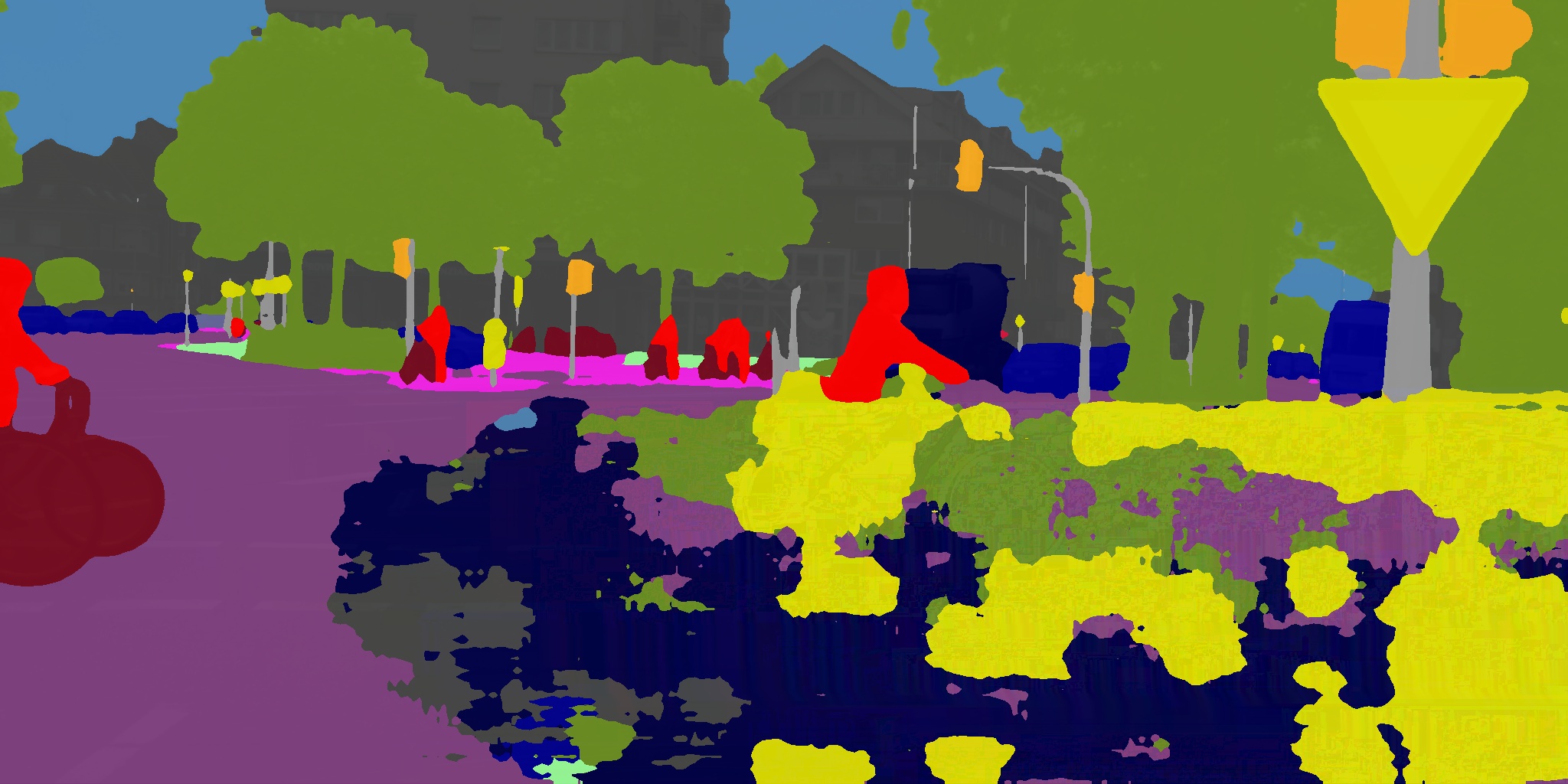}
		\end{minipage}\label{tra_19}
	} 
\hspace{0.15cm}
	\subfloat[Conventional 22 dB]{
		\begin{minipage}[b]{0.2\textwidth}
			\includegraphics[width=1.5in]{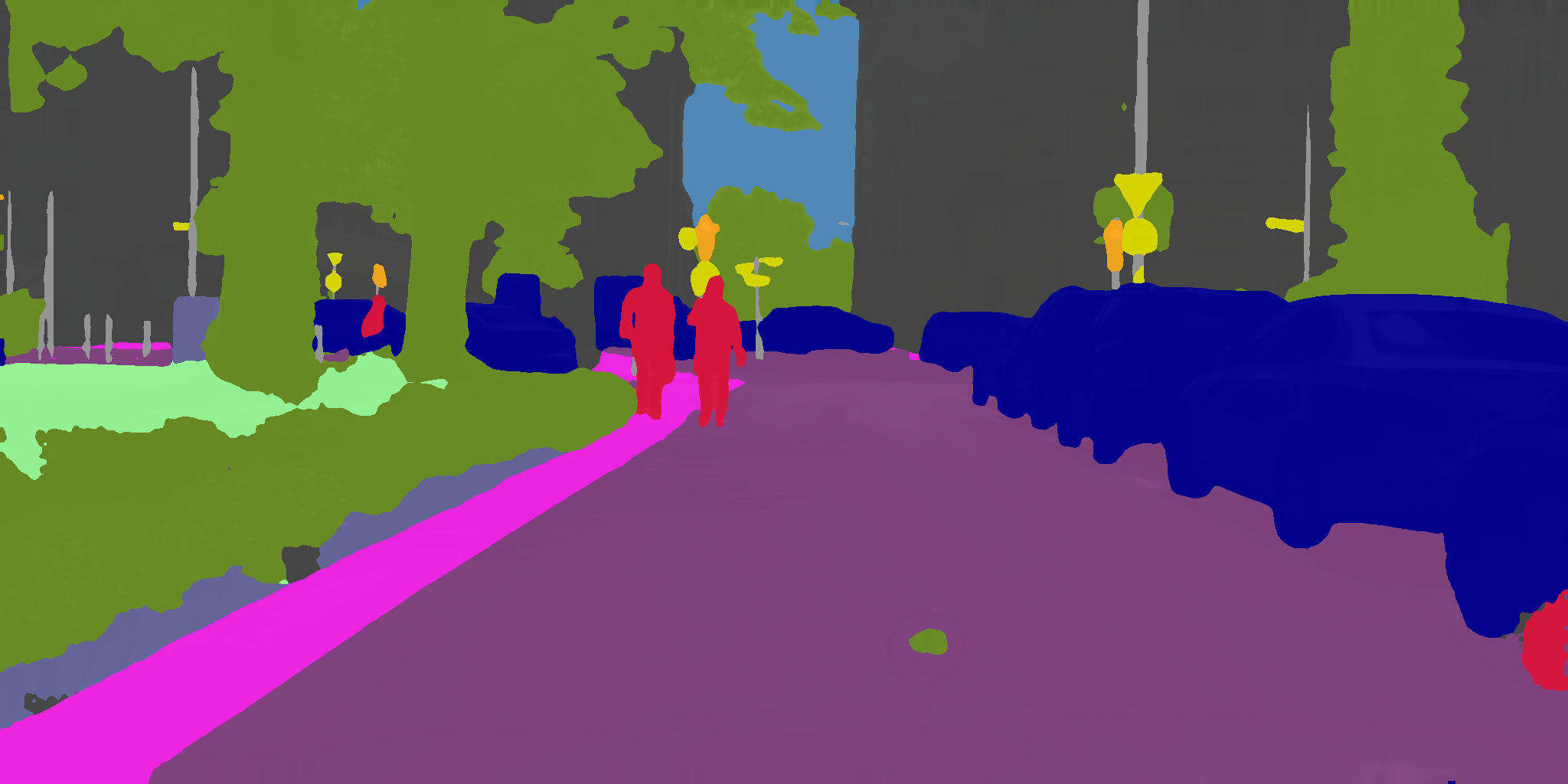}\\
			\vspace{-0.4cm}
			\includegraphics[width=1.5in]{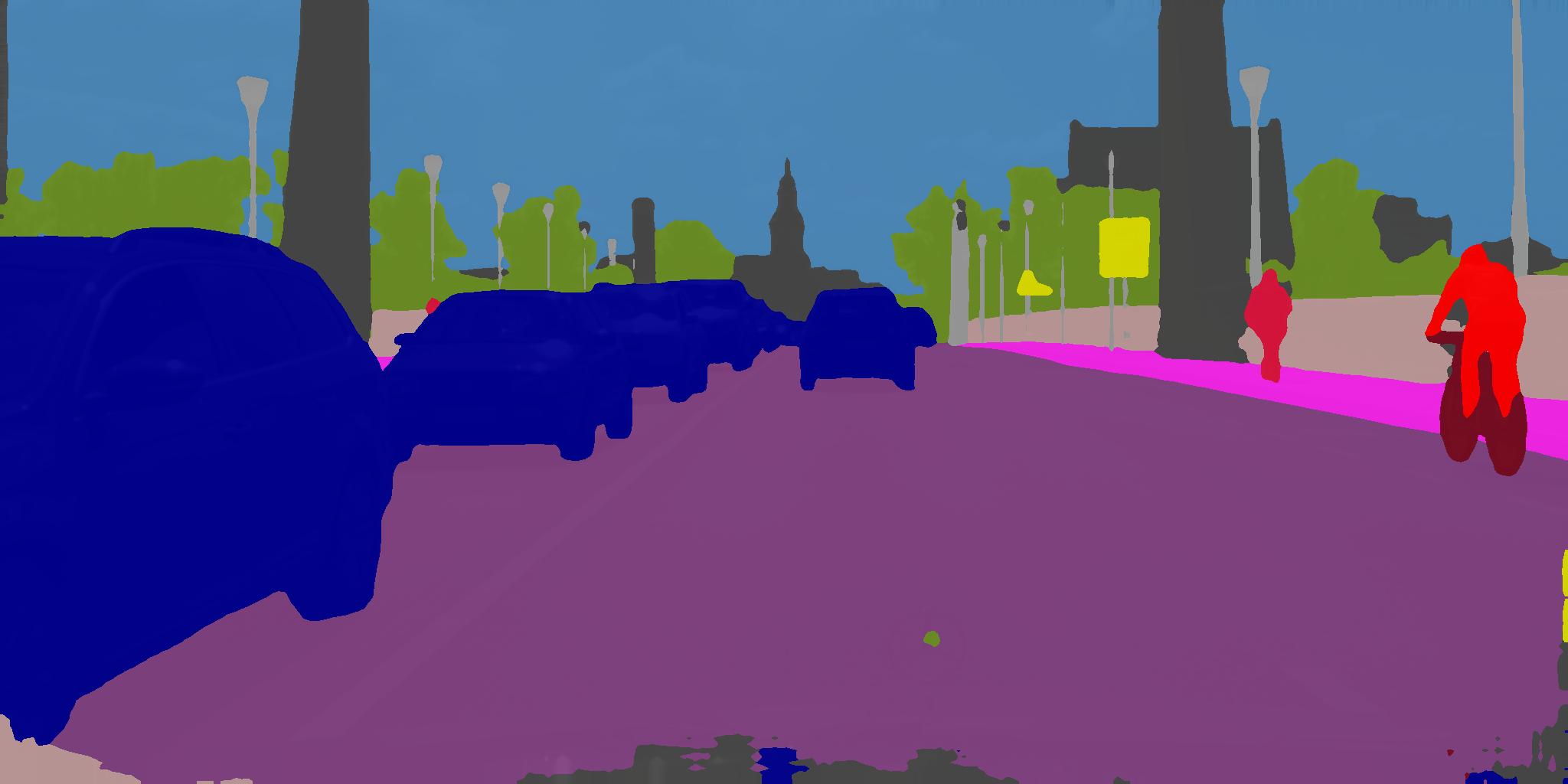}\\
			\vspace{-0.4cm}
			\includegraphics[width=1.5in]{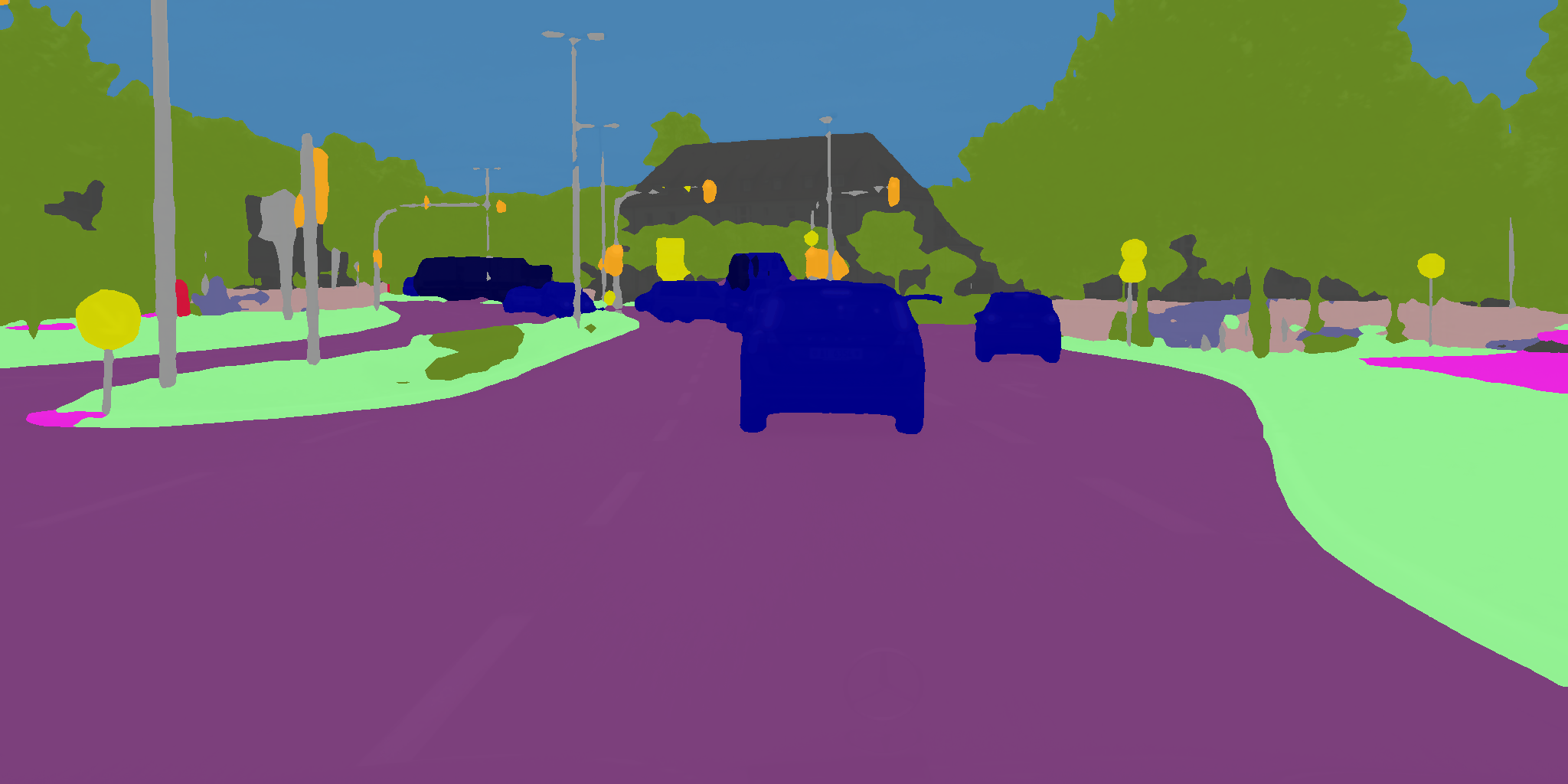}\\
			\vspace{-0.4cm}
			\includegraphics[width=1.5in]{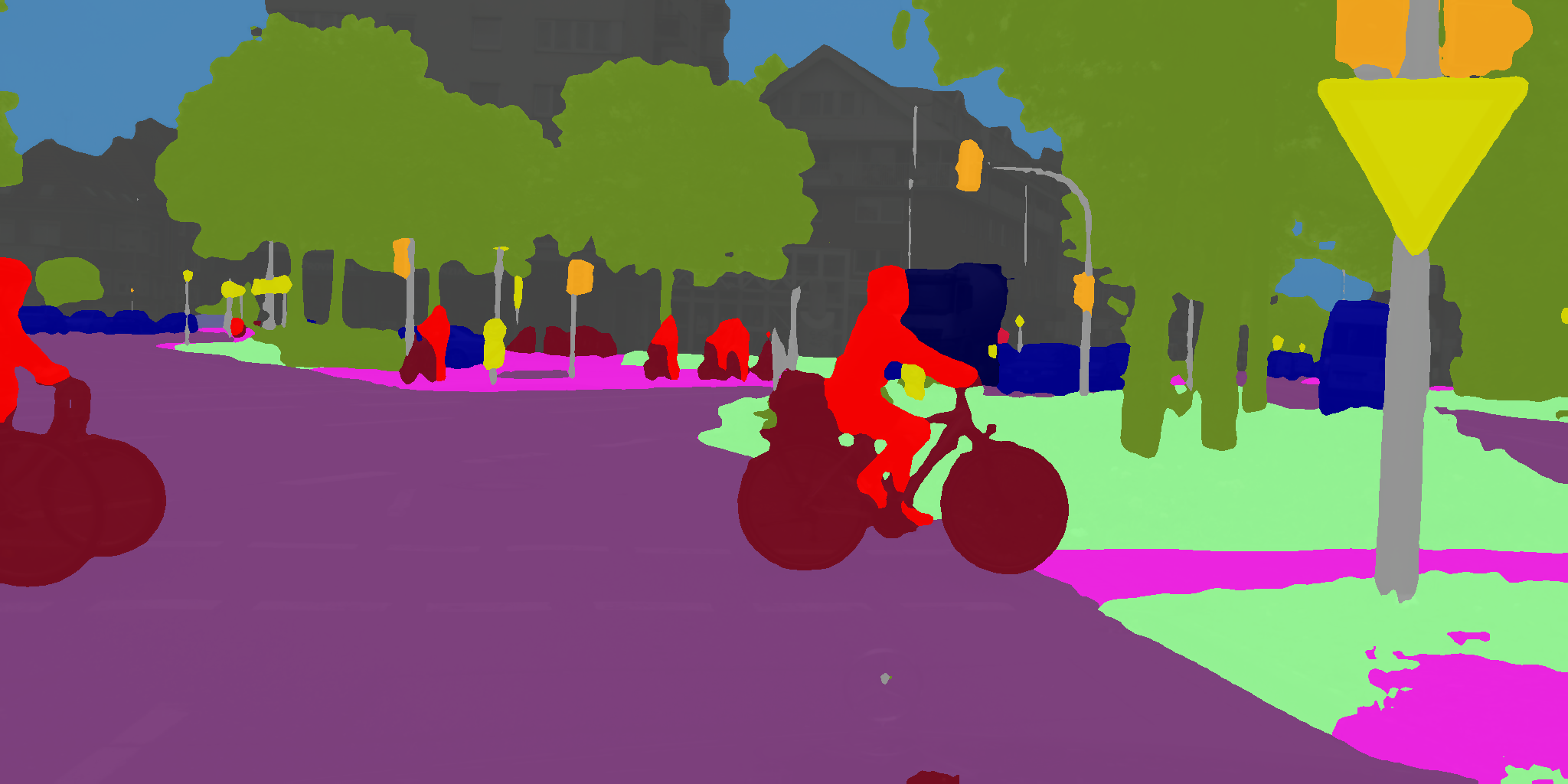}
		\end{minipage}\label{tra_22}
	} 
\hspace{0.15cm}
	\subfloat[VIS-SemCom 19 dB]{
		\begin{minipage}[b]{0.2\textwidth}
			\includegraphics[width=1.5in]{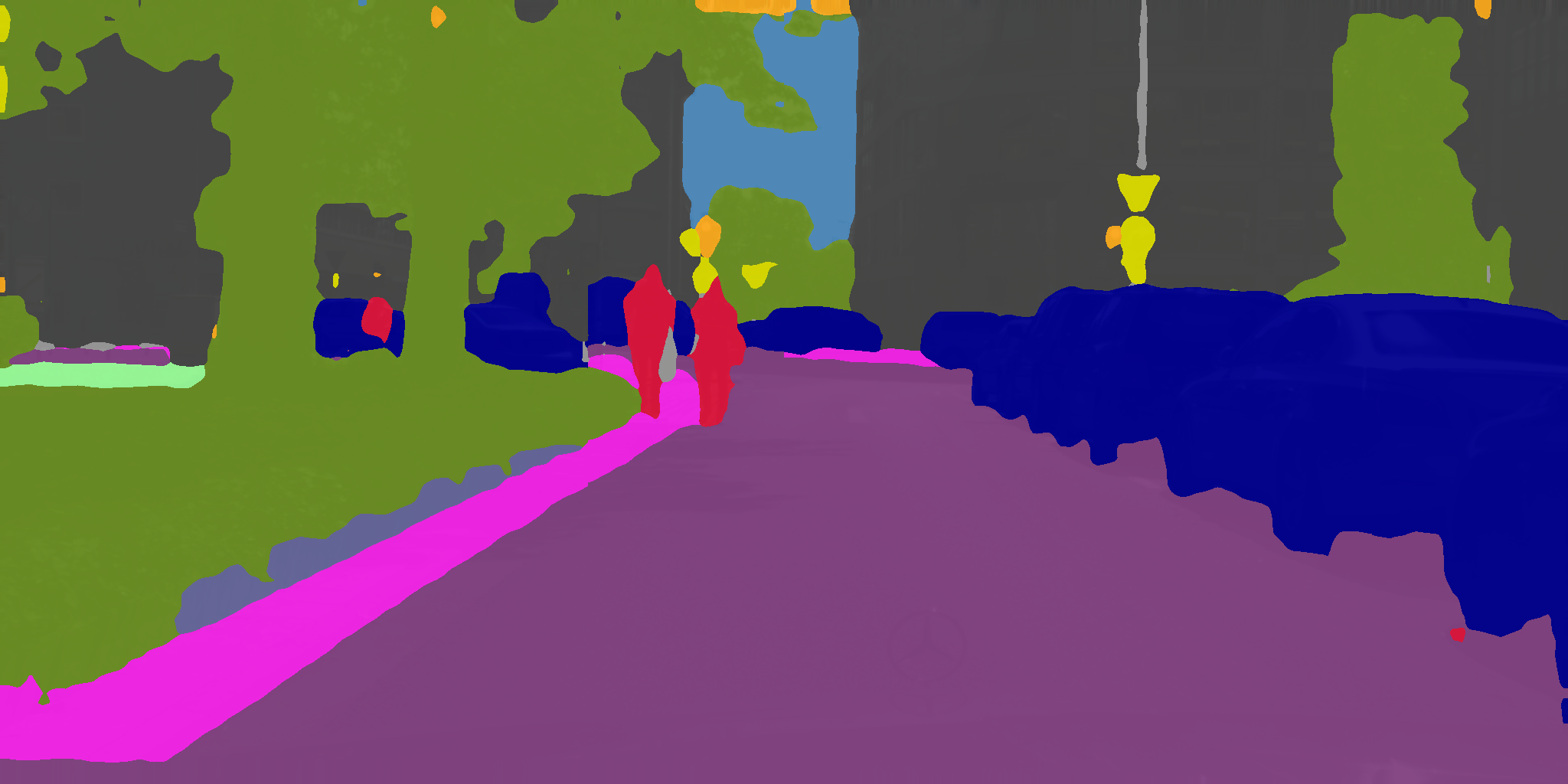}\\
			\vspace{-0.4cm}
			\includegraphics[width=1.5in]{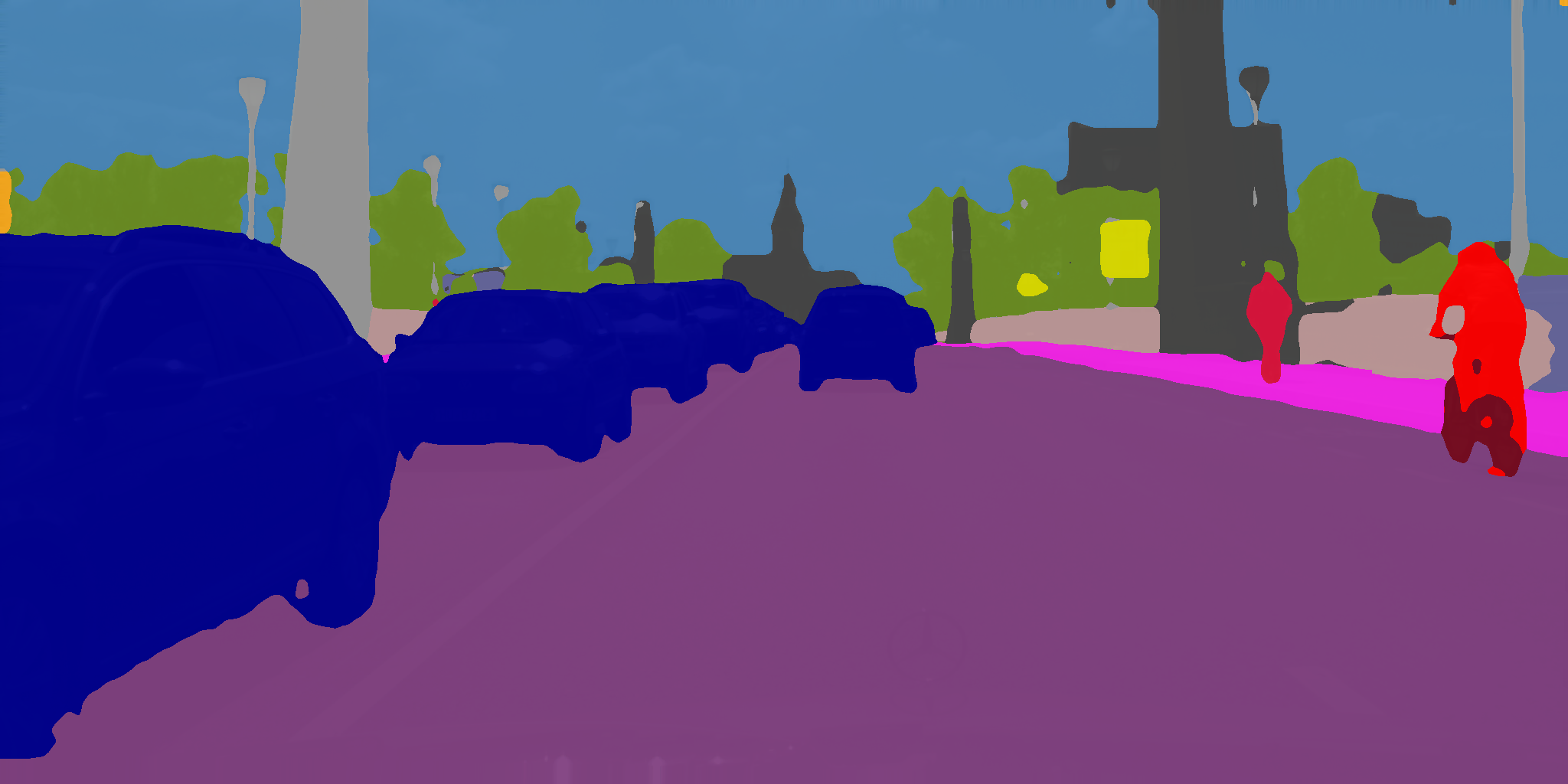}\\
			\vspace{-0.4cm}
			\includegraphics[width=1.5in]{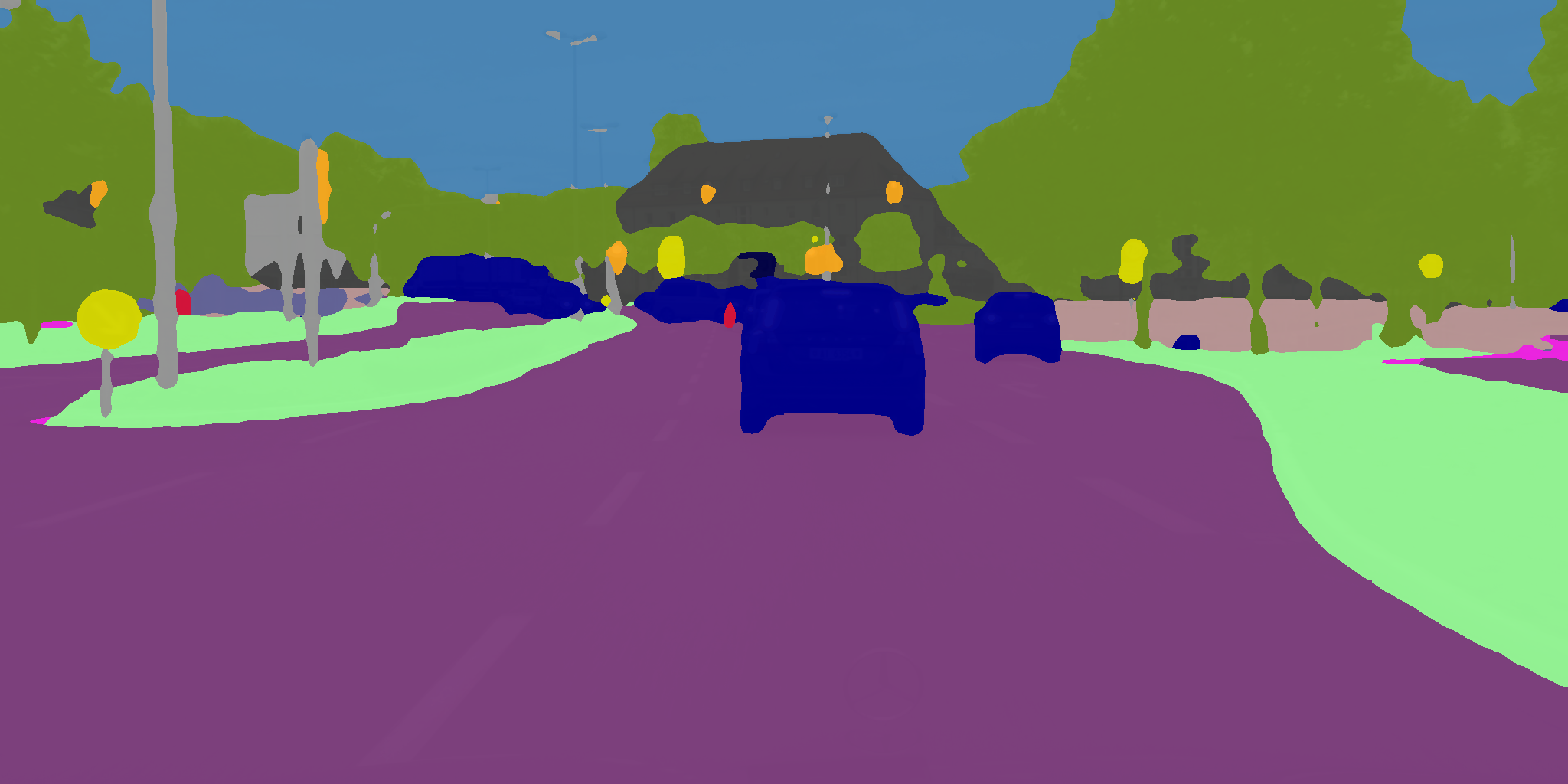}\\
			\vspace{-0.4cm}
			\includegraphics[width=1.5in]{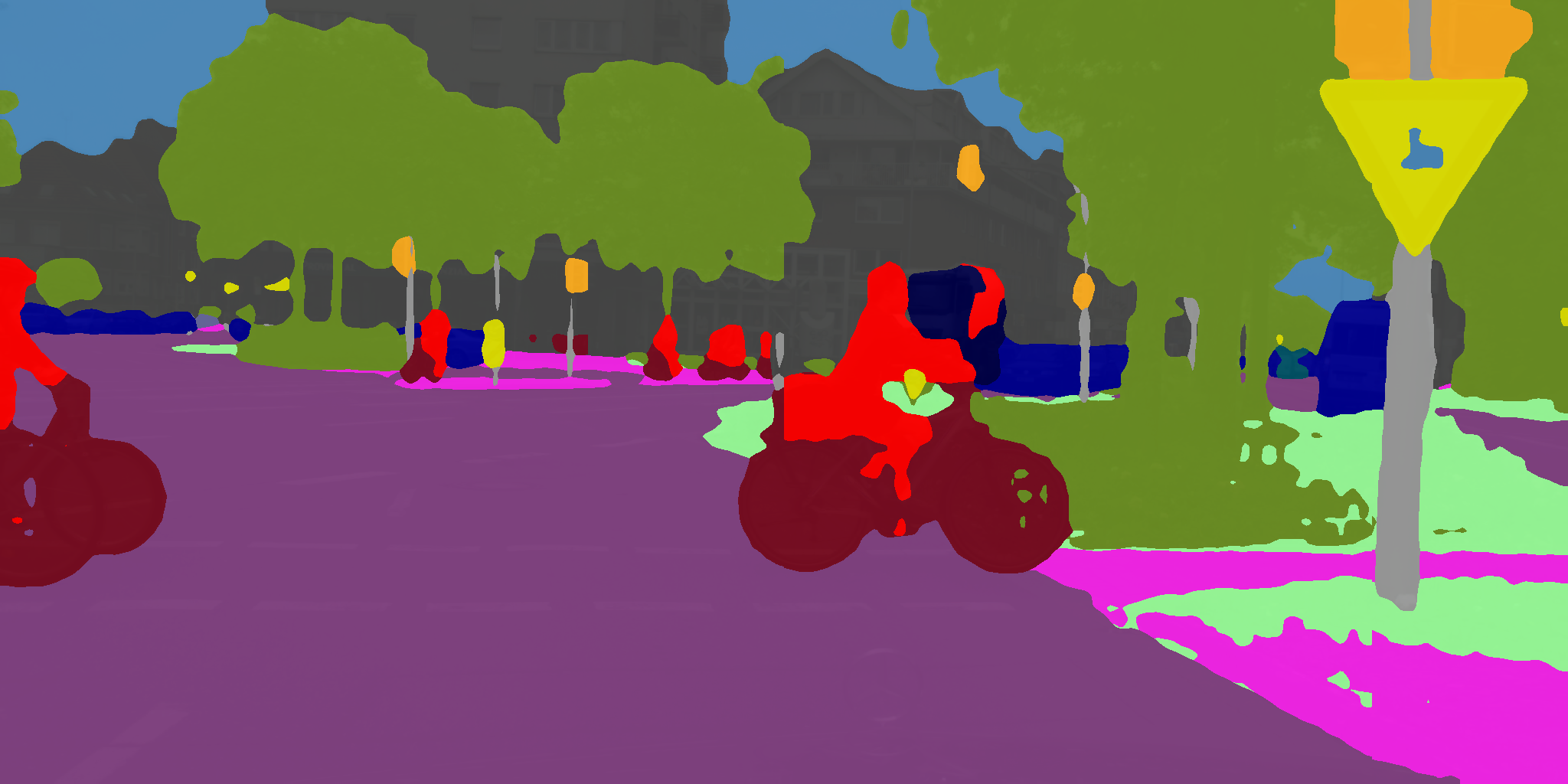}
		\end{minipage}\label{SC_19}
	}  
\hspace{0.15cm}
	\subfloat[VIS-SemCom 22 dB]{
		\begin{minipage}[b]{0.2\textwidth}
			\includegraphics[width=1.5in]{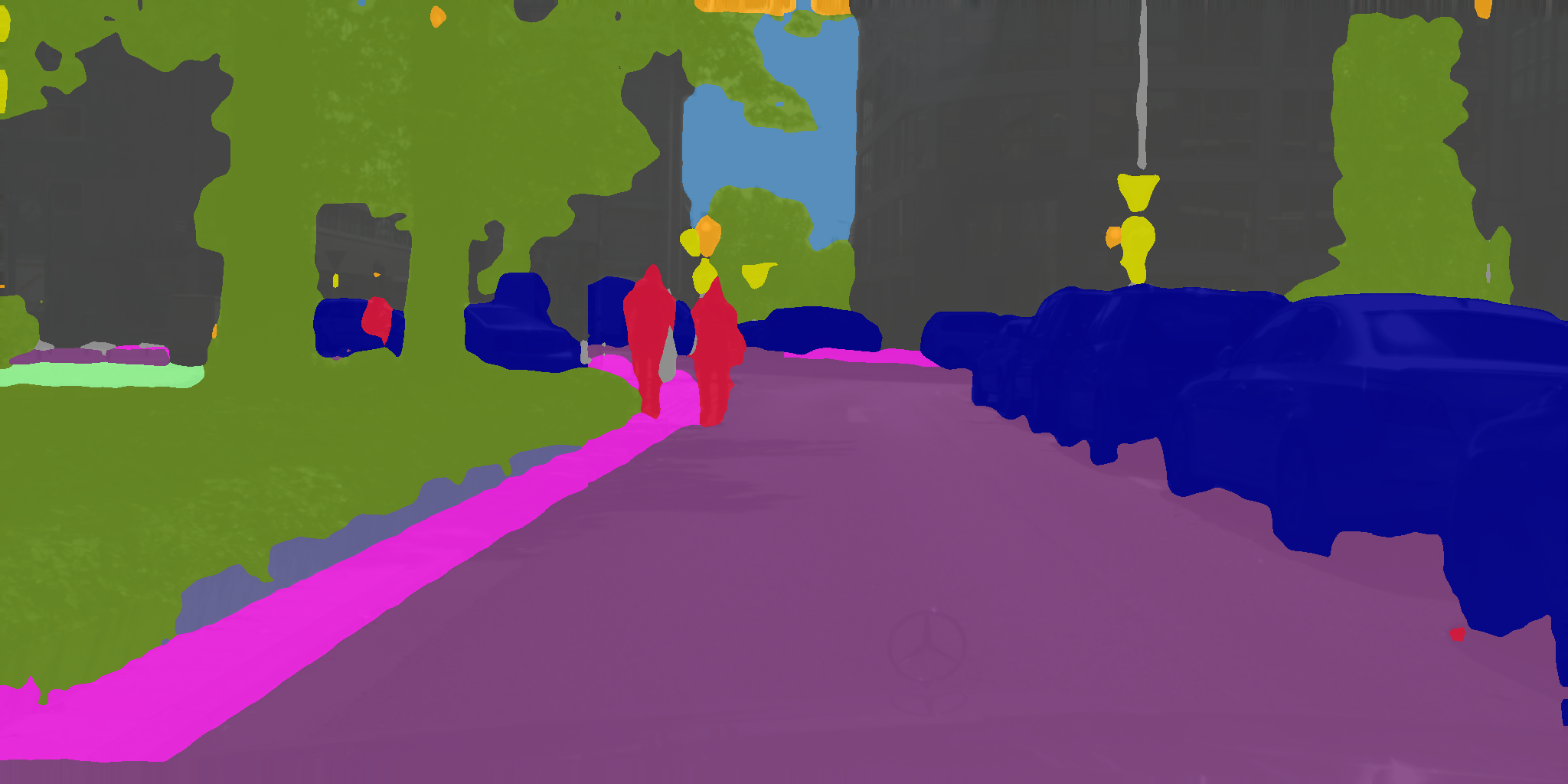}\\
			\vspace{-0.4cm}
			\includegraphics[width=1.5in]{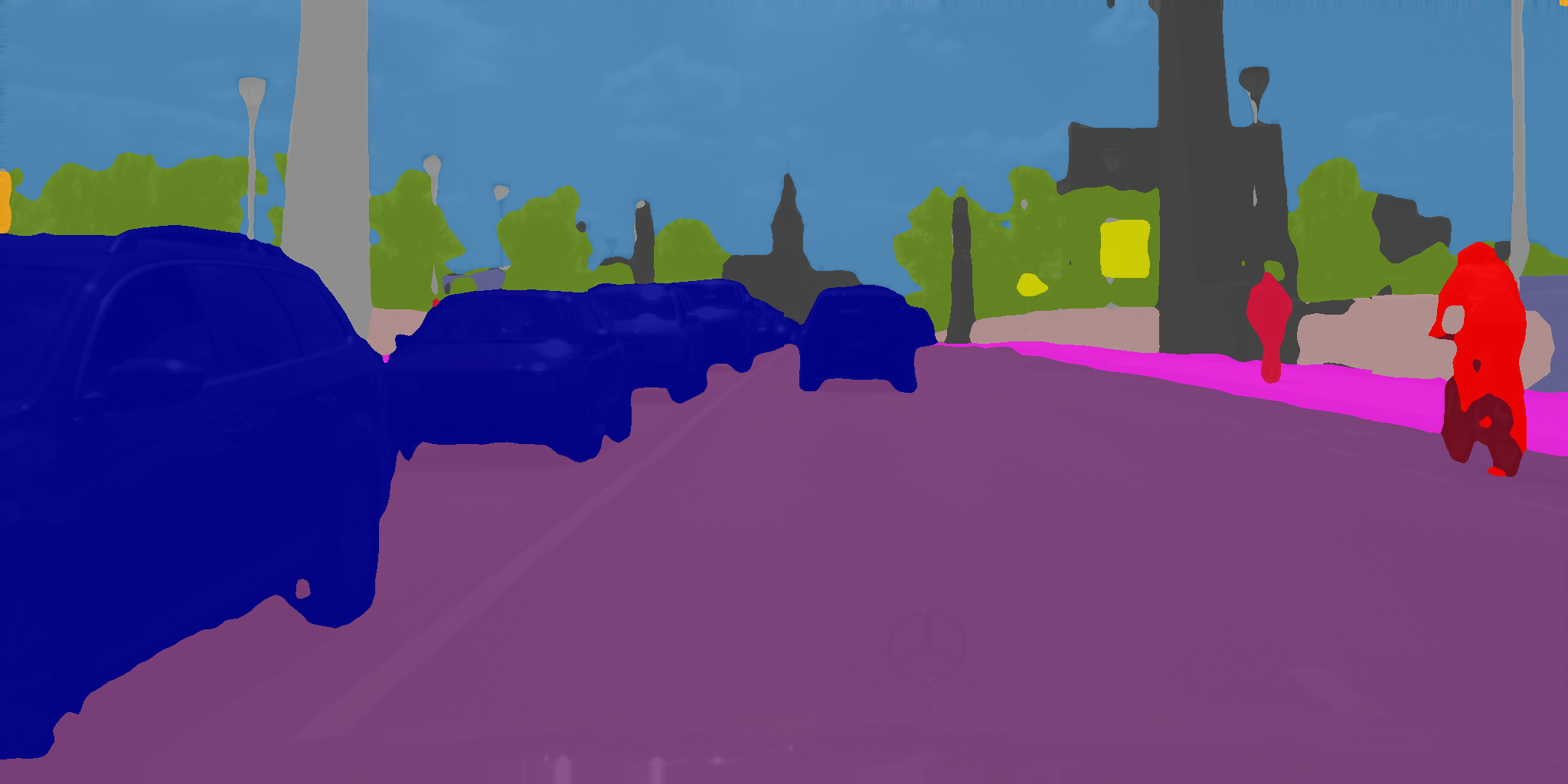}\\
			\vspace{-0.4cm}
			\includegraphics[width=1.5in]{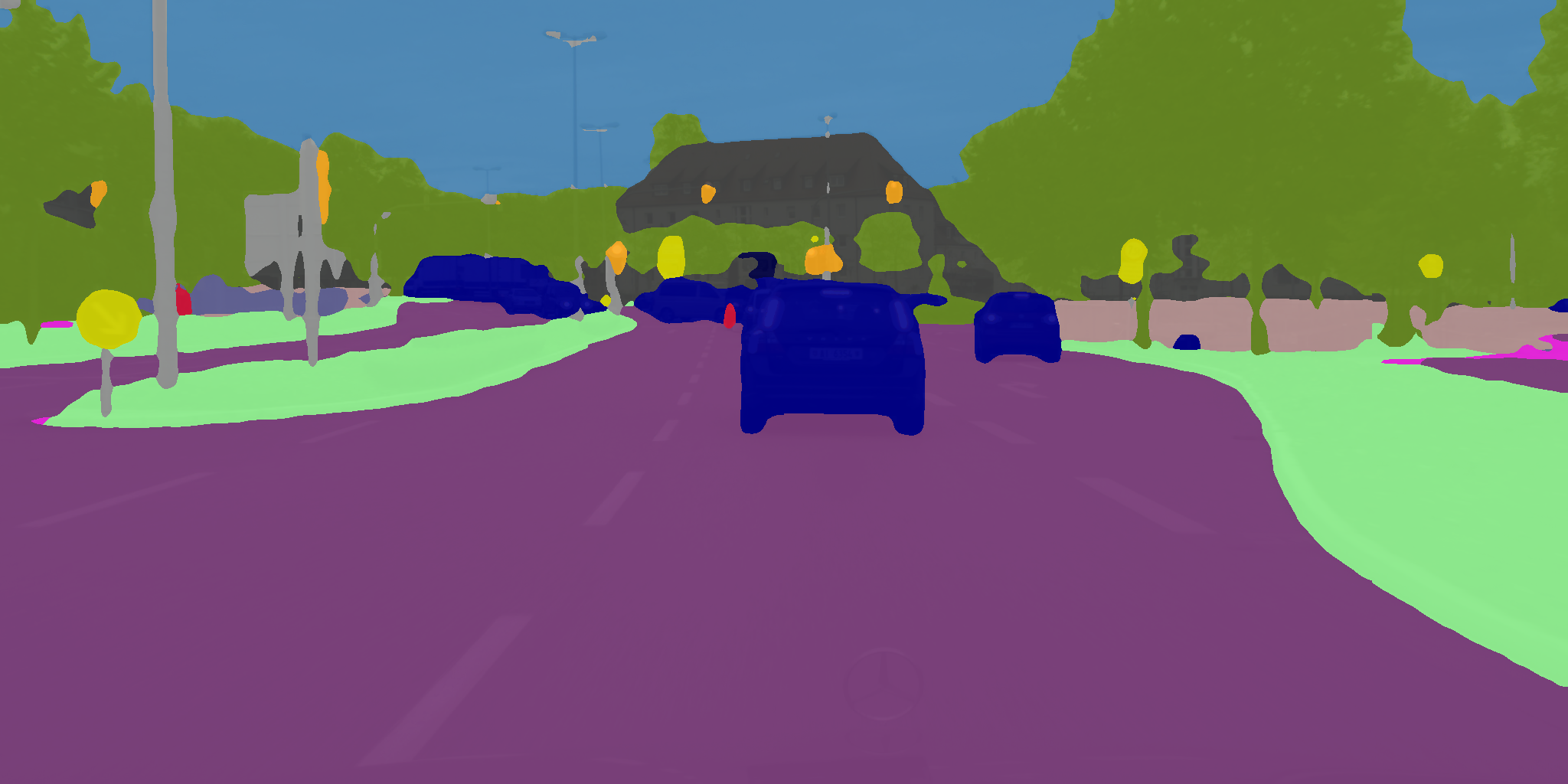}\\
			\vspace{-0.4cm}
			\includegraphics[width=1.5in]{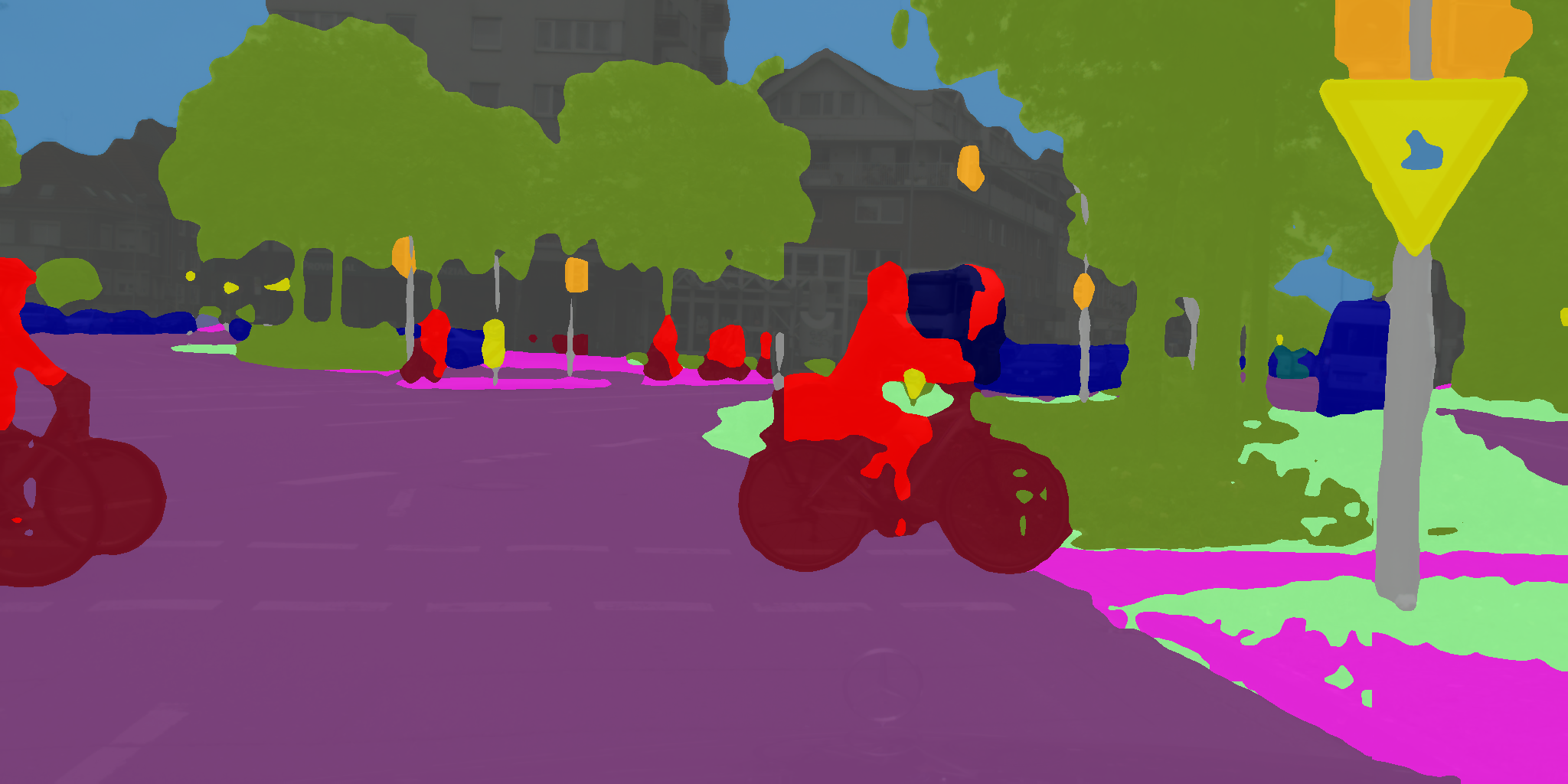}
		\end{minipage}\label{SC_22}
	}  
	\caption{The visualization results of image segmentation by the traditional scheme and the VIS-SemCom system with vehicle velocity 120 km/h and the compression ratio $R = 48$ (SNR$_{{test}}$ = 19 dB, 22 dB).}
	\label{Visual comparison}
\end{figure*}

We compare the proposed VIS-SemCom system with the conventional 
transmission scheme, which uses a joint photographic experts group (JPEG) format for source coding; low-density parity-check code (LDPC) with a 2/3 code rate as the channel coding; 4-QAM, 16-QAM and 64-QAM modulations; and DeepLabv3+~\cite{Chen2017} as well as OCRnet~\cite{Yuan2019} to segment the reconstructed original images thus realizing the image segmentation task at the receiver.

\subsection{MIoU Performance for The Whole Image}

Figure~\ref{Visual comparison} shows the visualization of image segmentation using the traditional scheme and VIS-SemCom system with vehicle velocity 120 km/h and compression ratio $R =48$. 
From Figure~\ref{Visual comparison}\subref{tra_19} and \subref{SC_19}, we can observe that the image label map recovered by the traditional scheme has a large distortion, while the proposed VIS-SemCom system can obtain high-quality image label maps. 
This is because VIS-SemCom only transmits semantic features with Swin Transformer backbone, and recovers the semantic segmentation results at the receiver, avoiding image decoding bit errors of the traditional scheme at low SNR values. 
From Figures~\ref{Visual comparison}\subref{tra_22} and \subref{SC_22}, we can see that, when SNR = 22 dB, VIS-SemCom can achieve segmentation performance comparable to the traditional scheme despite some detail loss, especially for high-correlation objects (e.g., cars, people, roads, etc.). 
This is due to the fact that the traditional scheme can fully recover the encoded image at high SNR conditions using more transmitted data.

\begin{figure}[t]
	\centering{\includegraphics[width=0.48\textwidth]{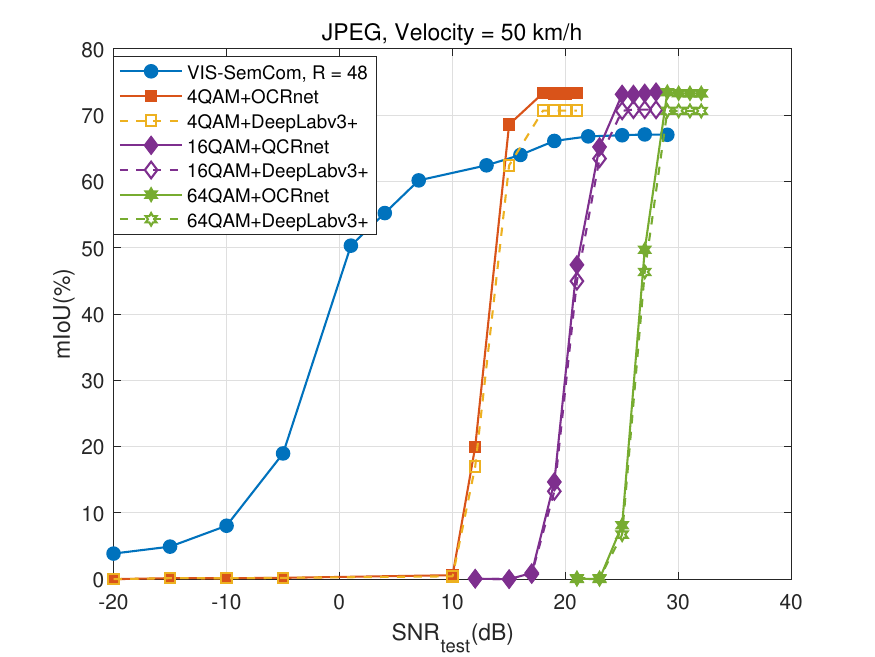}}
	\caption{The mIoU performance of the proposed VIS-SemCom scheme and traditional schemes with different sets. The vehicle velocity is 50 km/h.}
	\label{50km}
\end{figure}

\begin{figure}[t]
	\centering{\includegraphics[width=0.48\textwidth]{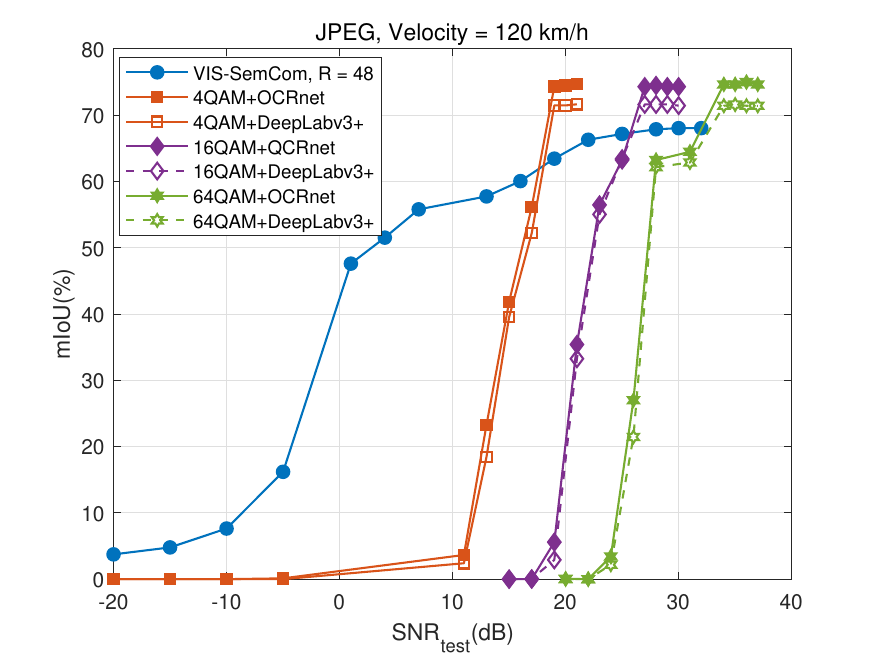}}
	\caption{The mIoU performance of the proposed VIS-SemCom scheme and traditional schemes with different sets. The vehicle velocity is 120 km/h.}
	\label{120km}
\end{figure}

Figure~\ref{50km} and Figure~\ref{120km} show how the mIoU changes as the SNR$_{test}$ increases, at vehicle velocities 50 km/h and 120 km/h, respectively. 
As expected, the mIoU of different schemes increases with SNR and our proposed VIS-SemCom outperforms traditional schemes under low SNR regimes. 
Compared with traditional schemes, the curves of the proposed VIS-SemCom are smooth, while traditional schemes suffer from a ``cliff effect"~\cite{Bourtsoulatze2019}, which is because of the bit errors in the original image at lower SNR, leading to a decrease in mIoU.
To describe the degree of performance improvement over the image transmission tasks after exploiting the VIS-SemCom system, the coding gain is defined as the difference (in dB) between SNRs required by different systems under a certain segmentation accuracy.
From Figure~\ref{50km}, we can observe that with the requirement of achieving 60\% segmentation mIoU, VIS-SemCom can obtain a coding gain of nearly 6 dB than the traditional JPEG + 4QAM + OCRnet, and a 2 dB coding gain can be achieved under the high-speed condition in Figure~\ref{120km}.
\emph{ The main reason for the above phenomenons is that VIS-SemCom can learn and adapt to channel changes and achieve semantic-level error correction. }
In addition, VIS-SemCom has a lower mIoU in high SNR ranges than the traditional schemes because traditional schemes can reconstruct the complete image thus obtaining better segmentation results.
Comparing Figures~\ref{50km} and \ref{120km}, we can also find that the higher vehicle velocity leads to a decrease in the mIoU performance of VIS-SemCom and traditional schemes. The performance is because the Doppler effect under dynamic channels affects the signal demodulation and the accuracy of image segmentation. 

\begin{figure}[t]
	\centering{\includegraphics[width=0.48\textwidth]{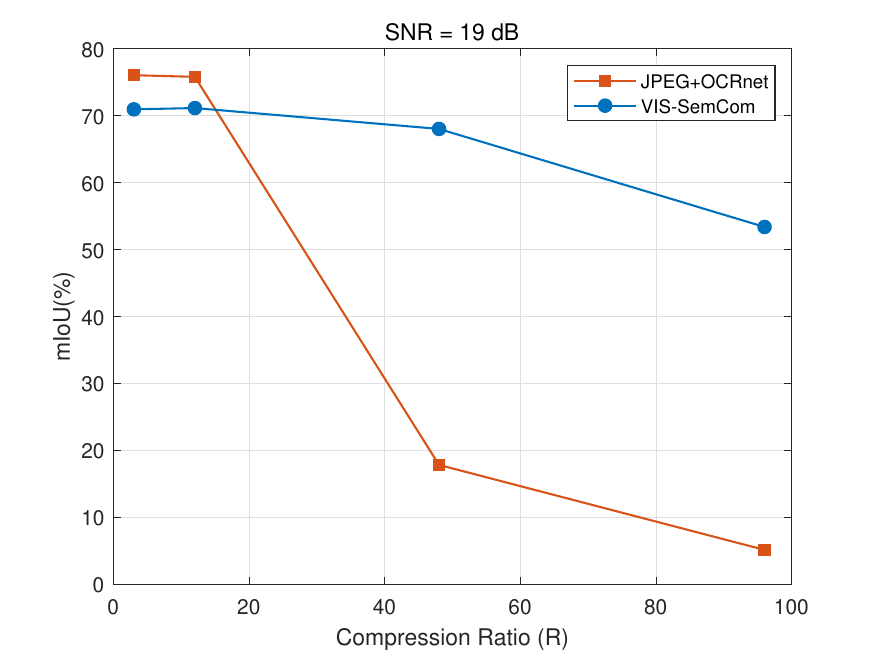}}
	\caption{The mIoU performance of the proposed VIS-SemCom scheme and traditional JPEG+OCRnet method with different compression ratios at SNR = 19 dB.}
	\label{sem_tra}
\end{figure}

In Figure~\ref{sem_tra}, we further compare the mIoU with various compression ratios of the proposed VIS-SemCom and JPEG + OCRnet scheme at 19 dB SNR.
It can be observed that the mIoU of the two methods decreases along with the increase of the compression ratio. 
Moreover, the mIoU of VIS-SemCom decreases smoothly, while that of JPEG + OCRnet drops abruptly when the compression ratio $R$ exceeds 12.
This phenomenon proves that VIS-SemCom can maintain high segmentation accuracy with a small amount of transmitted data. 
When meeting a 60\% segmentation mIoU requirement of autonomous driving, VIS-SemCom can reduce the data amount by about 70\% compared to JPEG + OCRnet.
The phenomenon is because JPEG + OCRnet eliminates high-frequency signals, resulting in loss and blurring of important details for image segmentation, meanwhile, mass artifacts and noise lead to local discontinuity and non-uniformity in images. 
In contrast, the VIS-SemCom can extract and aggregate multi-scale semantic information of images with Swin Transformer retaining the most key information, and reducing the channel impact via extracting contextual correlation.

\begin{figure}[t]
	\centering{\includegraphics[width=0.48\textwidth]{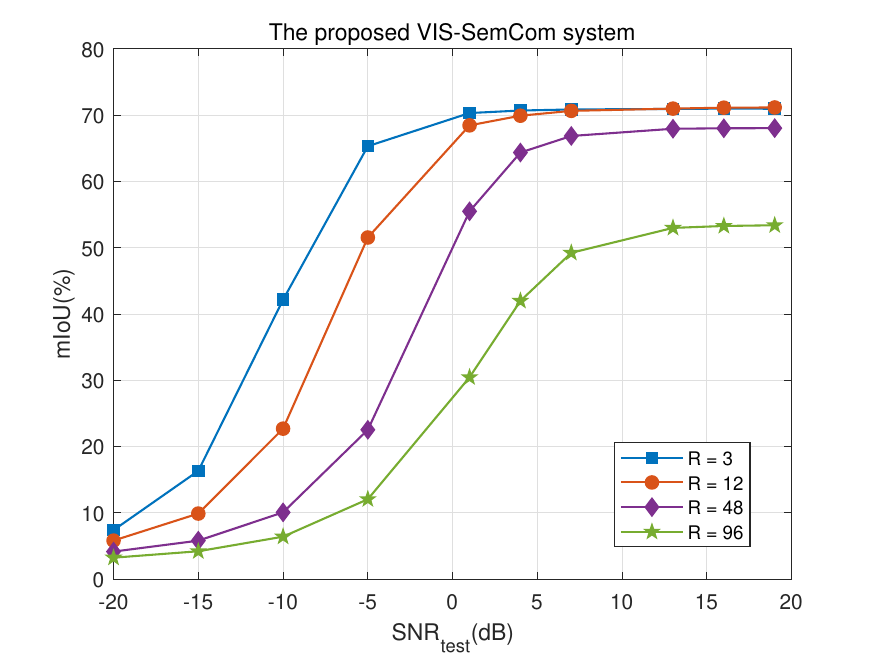}}
	\caption{The mIoU performance of the proposed VIS-SemCom scheme in different compression ratios $R$ with respect to SNR values.}
	\label{sem_R}
\end{figure}

Figure~\ref{sem_R} presents the mIoU of the proposed VIS-SemCom in different compression ratios versus different SNR values. 
From Figure~\ref{sem_R} we see that, for all compression ratios, the mIoU increases as the SNR value increases, and then tends to saturation. 
In particular, when the SNR is 5 dB, the mIoU can reach nearly 70$\%$ when $R = 3, 12, 48$,  which meets the demands of autonomous driving. 
Additionally, the segmentation mIoU decreases as the compression ratio increases due to the less transmitted information, which is consistent with the trend in Figure~\ref{sem_tra}.

\subsection{IoU Performance for Different Objects}

\begin{table*}[t]
	\centering
	\caption{The IoU (\%) performance of different objects. Compared with a benchmark (the last column), results with substantial improvement are in bold. Besides, compared with the results in the third column, results with an improvement are underlined.}
	\begin{tabular}{|c|c|c|c|c|}
		\hline
		Classes  & \makecell{Weight coefficient \\setting $w_i$}  &\makecell{ Loss using Eq.~\eqref{total_loss}} & \makecell{ With OHEM strategy } & \makecell{Loss using traditional cross entropy} \\
		\hline
		\textbf{road} & 1.25595 & \textbf{97.85}   & 97.52  & 97.56 \\
		\hline
		\textbf{sidewalk} & 1.377 & \textbf{82.71}   & \underline{\textbf{84.87}} & 81.20 \\
		\hline
		\textbf{building} & 1.299 & \textbf{93.40}    & \underline{\textbf{93.96}} & 91.05 \\
		\hline
		\textbf{vegetation} & 1.3179 & \textbf{93.25}   & 92.31 & 92.35 \\
		\hline
		\textbf{ person} & 1.47645 & \textbf{79.67} & \underline{\textbf{81.40}} & 77.20  \\
		\hline
		\textbf{rider} & 1.6674 & \textbf{57.50} &\underline{\textbf{62.60} }& 56.49  \\
		\hline
		\textbf{cars} & 1.35555 & \textbf{93.58}  & \textbf{93.24} & 92.64  \\
		\hline
		\textbf{truck} & 1.62975 & \textbf{48.85} & \underline{\textbf{52.54}} & 48.50 \\
		\hline
		\textbf{bus} & 1.64325 & \textbf{83.45} & \underline{\textbf{85.78}}  & 81.67 \\
		\hline
		\textbf{train} & 1.62975 & \textbf{78.35} & \underline{\textbf{82.40}}  & 76.20 \\
		\hline
		\textbf{motorcycle} & 1.72935 & \textbf{62.40} &\underline{\textbf{ 66.43}}  & 61.30 \\
		\hline
		\textbf{ bicycle} & 1.57605 & 70.96 & \underline{\textbf{74.69}} & 71.40  \\
		\hline
	\end{tabular}
	\label{table3}%
\end{table*}%

\begin{table}[t]
	\centering
	\caption{The IoU performance of systems with different STB combinations on important objects. Results with substantial improvement are in bold.}
	\begin{tabular}{|c|c|c|c|c|}
		\hline
		\multirow{2}{*}{\textbf{Classes}}
		& \multicolumn{4}{c|}{\textbf{STB combinations}} \\
		\cline{2-5}
		& [3, 5, 7, 9] & [2, 2, 2, 18] & [2, 2, 9, 9] & [2, 2, 18, 2] \\
		\hline
		\textbf{road} & 96.45\% & 95.67\% & \textbf{97.56\%} & 97.54\% \\
		\hline
		\textbf{sidewalk} & 75.65\% & 71.43\% & \textbf{81.20\%} & 80.99\% \\
		\hline
		\textbf{building} & 88.56\% & 86.35\% & \textbf{91.05\%} & 90.88\% \\
		\hline
		\textbf{vegetation} & 89.21\% & 89.23\% & \textbf{92.35\%} & 91.36\% \\
		\hline
		\textbf{person} & 72.30\% & 73.50\% & \textbf{77.20\%} & 76.92\% \\
		\hline
		\textbf{rider} & 54.35\% & 52.30\% & 56.49\% & 57.55\% \\
		\hline
		\textbf{cars} & 87.41\% & 88.48\% & 92.64\% & 92.68\% \\
		\hline
		\textbf{truck} & 44.30\% & 45.60\% & \textbf{48.50\%} & 47.82\% \\
		\hline
		\textbf{bus} & 75.55\% & 74.20\% & \textbf{81.67\%} & 81.59\% \\
		\hline
		\textbf{train} & 74.30\% & 73.50\% & \textbf{76.20\%} & 75.28\% \\
		\hline
		\textbf{motorcycle} & 55.25\% & 54.23\% & \textcolor{red}{\textbf{61.23\%}} & 57.76\% \\
		\hline
		\textbf{bicycle} & 68.00\% & 63.47\% & 71.40\% & 72.40\% \\
		\hline
	\end{tabular}
	\label{table2}%
\end{table}%


To provide a comprehensive perspective of the proposed VIS-SemCom, we present the IoU values for 12 classes of important objects in the safety of autonomous driving in Tables~\ref{table2} and \ref{table3}, including road, sidewalk, building, vegetation, person, rider, car, truck, bus, trains, motorcycle, and bicycle.

Table~\ref{table2} shows the IoU for each object under different STB combinations. 
During the test stage, we set the total number of STBs as 24, and change the STB into 4 different combinations which are [3, 5, 7, 9], [2, 2, 2, 18], [2, 2, 9, 9], and [2, 2, 18, 2], respectively.
The other network parameters are the same as Table~\ref{table1}, the compression ratio $R = 3$, SNR$_{test}=19$ dB, and the number of training iterations is 160000. 
Based on the combination [2, 2, 18, 2], the networks of the other three combinations are obtained by transfer learning, which speeds up the convergence of training. 
In Table~\ref{table2}, it can be observed that the combination [2, 2, 9, 9] has the best segmentation accuracy of each class and achieves substantial classification accuracy improvement. 
For instance, the improvement of IoU of motorcycle can be up to $3.5\%$, the accurate identification of which is considerably important in autonomous driving.  
The improvement is because more STBs on the low-resolution feature map are allocated, resulting in balancing the semantic extraction of the latter two stages and increasing the global perceptual field. 
Table~\ref{table2} demonstrates that the proposed multi-scale semantic extractor based on Swin Transformer can enhance the accuracy of important objects. 

Table~\ref{table3} shows the performance of the designed training loss Eq.~\eqref{total_loss} and OHEM training strategy in Section~\ref{training_strategy}.
In the simulation, the weight coefficient for each class is obtained by multiplying the category balance loss coefficient and the attention coefficient. The category balance loss coefficient is set according to~\cite{Chen2017}. The attention coefficients for the 12 important object classes are set to 1.5, and those of the other classes are set to 1.
In this setting, we can observe the improvements of the IoU for nearly all classes compared with the NN model using the traditional cross-entropy loss function. The outperformance is because the designed loss assigns weights to different objects thus enhancing the performance of important objects.
From Table~\ref{table3}, we can also see that the effect of some objects has a low increase of less than $0.5\%$, such as motorcycle and truck, or even a small decrease, such as bicycle. 
These objects are hard samples and difficult to learn.
The fourth column of Table~\ref{table3} shows the proposed OHEM strategy performance, in which $thresh = 0.7$ and $min\_kept = 100000$.
After using the OHEM training strategy, we can observe an improvement of about $4\%$ for the hard sample objects, such as motorcycle and bicycle. 
Meanwhile, a small improvement for the objects with good segmentation accuracy~(sidewalk, building, etc.), and even a slight decrease~(car, road, etc.). 
This is because that the OHEM strategy focuses on training object pixels with lower confidence and less proportion. 
Table~\ref{table3} verifies the effectiveness of network training loss and the designed strategy for enhancing the performance of important objects and small sample learning.

\section{CONCLUSION}
In this article, we have proposed a VIS-SemCom system built on the Swin Transformer backbone to perform image segmentation tasks, achieving an understanding of important objects in autonomous driving scenes. Specifically, we design a multi-scale semantic feature extractor to mine global semantic information of images, and transmit the semantic feature representation as the compressed image data into wireless channels. To train the proposed VIS-SemCom, a novel loss function combining weighted multi-class cross-entropy and IoU is considered. Furthermore, an OHME training strategy is utilized to deal with a relatively small number of samples in the data set. Finally, we conducted various experiments across a widely recognized autonomous driving dataset. Results have demonstrated that the proposed VIS-SemCom system can transmit a small amount of data to obtain the same segmentation accuracy as traditional schemes. In addition, the segmentation accuracy of important objects is significantly improved, which is in line with the requirements of autonomous driving safety. 

This work is a preliminary exploration of applying semantic communication to vehicular networks for reliable autonomous driving. Our future research will delve into the complexities of the vehicular environment.



 \def\baselinestretch{0.8}
 \bibliographystyle{IEEEtran}
 \bibliography{IEEEabrv,myref}

\end{document}